\documentclass[12pt]{article}
\usepackage{cite}
\usepackage{float}
\usepackage{amsfonts}
\usepackage{amsmath}
\usepackage{amsbsy}
\usepackage{graphicx}
\usepackage{amssymb}
\usepackage{amsthm}
\usepackage{bm}
\usepackage{epsfig}
\usepackage{latexsym}
\usepackage{pdflscape}
\usepackage{color}
\numberwithin{equation}{section}
\usepackage{young}
\allowdisplaybreaks

\DeclareMathOperator{\tr}{tr}
\DeclareMathOperator{\Tr}{Tr}
\DeclareMathOperator{\Ai}{Ai}

\DeclareMathOperator{\diag}{diag}

\def\cO{\mathcal{O}}

\def\({\left(}
\def\){\right)}

\newcommand{\pd}{\partial}

\newcounter{aff}

\begin{document}

\begin{titlepage}
\begin{flushright}
{\footnotesize DESY 15-101, OCU-PHYS 428}
\end{flushright}
\begin{center}
{\Large\bf Exact Instanton Expansion of ABJM Partition Function}

\bigskip\bigskip
{\large 
Yasuyuki Hatsuda\footnote[1]{\tt yasuyuki.hatsuda@desy.de},
\quad
Sanefumi Moriyama\footnote[2]{\tt moriyama@sci.osaka-cu.ac.jp},
\quad
Kazumi Okuyama\footnote[3]{\tt okuyama@azusa.shinshu-u.ac.jp}
}\\
\bigskip
${}^{*}$\,
{\small\it DESY Theory Group, DESY Hamburg,\\
Notkestrasse 85, D-22603 Hamburg, Germany}
\medskip\\
${}^{\dagger}$\,
{\small\it 
Department of Physics, Osaka City University,\\
Osaka 558-8585, Japan}
\medskip\\
${}^{\ddagger}$\,
{\small\it Department of Physics, Shinshu University\\
Matsumoto 390-8621, Japan}
\end{center}

\begin{abstract}
We review recent progress in determining the partition function of the ABJM theory in the large $N$ expansion, including all of the perturbative and non-perturbative corrections.
Especially, we will focus on how these exact expansions are obtained from various beautiful relations to Fermi gas system, topological string theory, integrable model and supergroup.
\end{abstract}

\end{titlepage}
\tableofcontents

\section{Introduction}\label{intro}
Chern-Simons theories play a central role in modern string theory.
Two major subjects in early days are open string field theory \cite{sft} and topological string theory \cite{cs}.
It was known not only that Chern-Simons theory satisfies the axiom of topological field theory but also that the topological theory can be lifted to topological string theory.
Also, the beautiful gauge invariant structure was used to construct the covariant open string field theory.
Both of these developments are deeply related to understanding non-perturbative aspects of string theory.

The relation between the Chern-Simons theory and the non-perturbative string theory becomes even more important in the supersymmetrization.
It was known that up to ${\cal N}=3$ the supersymmetric Chern-Simons theory can be constructed for any gauge group and any representation \cite{ZK,KL,KLL}.
Also, it was noted that when the levels are summed up to zero, the theory enjoys the conformal symmetry \cite{GY,JT}.
After some special arguments of the ${\cal N}=4$ enhancements \cite{GW,HLLLP1}, finally it was found that when the gauge group is $U(N)_k\times U(N)_{-k}$ with the subscripts denoting levels $k$ and $-k$ and the matters are in the bifundamental representation, the supersymmetry is enhanced to ${\cal N}=6$ \cite{ABJM}, which is called the ABJM theory.
Later, it was also found \cite{HLLLP2,ABJ} that even when the ranks of two factors in the gauge group are different, $U(N_1)_k\times U(N_2)_{-k}$, the Chern-Simons theory still enjoys the enhanced supersymmetry ${\cal N}=6$, which is called the ABJ theory.
So we shall often refer to them inclusively as ABJ(M) theory.

The discovery of this maximal supersymmetric conformal Chern-Simons theory enables us to open up another window to non-perturbative aspects of string theory.
In the study of the non-perturbative effects in string theory, we were naturally led to a conjecture that there exists an eleven-dimensional theory, which is dubbed M-theory \cite{11D}.
It is expected that this theory is the {\it mother} theory with fundamental {\it membrane} excitation from which our perturbative string theory stems after compactifications, though only little is known about this {\it mysterious} theory.
From the supergravity analysis, it is known that, besides the fundamental electric M2-branes, there exist solitonic magnetic M5-branes.
Also, from the AdS/CFT correspondence, we know that the degrees of freedom of $N$ M2-branes and $N$ M5-branes are $N^{3/2}$ and $N^3$ respectively \cite{KT}.

The ABJM theory describes a stack of $N$ multiple M2-branes put on the target space ${\mathbb C}^4/{\mathbb Z}_k$, while the ABJ theory describes a composite system of both $N_1$ multiple M2-branes and $N_2-N_1$ multiple fractional M2-branes for $N_2>N_1$.
For other superconformal Chern-Simons theory with less supersymmetries \cite{GY,JT,BKKS}, we expect that they describe multiple M2-branes on various backgrounds with less supersymmetries.
It is then natural to ask whether we can reproduce the degrees of freedom by studying the free energy of the ABJM theory in the large $N$ limit.
Also, it is interesting to see how the large $N$ corrections behave and whether there is any interesting physical implication in it.

In contrast to Chern-Simons theory, let us shortly comment on Yang-Mills theory.
Along with the Chern-Simons theory, Yang-Mills theory also plays a crucial role in modern string theory.
Especially, it is known that multiple D-branes are described by the maximally supersymmetric Yang-Mills theory with the gauge group $U(N)$ \cite{ym}.
This was found by identifying the dynamical degrees of freedom of D-branes as open string excitations appearing in each pair of the multiple D-branes.
The free energy $N^2$ in the large $N$ limit is interpreted as the number of the matrix elements and matches with the supergravity computation \cite{KT}.

Although the maximally supersymmetric Chern-Simons theory selects a special gauge group and a special representation, the maximally supersymmetric Yang-Mills theory is defined for general gauge groups and representations.
Then, it is natural to expect that the gauge group and the matter contents of the ABJ(M) theory are special.
In fact it was observed in \cite{DT} that, when we look at the matrix model appearing after applying the localization techniques, the gauge group and the matter content have a good interpretation in the supergroup $U(N_1|N_2)$.
Interestingly, the relation to the supergroup $U(N_1|N_2)$ persists even in the one-point function of the half-BPS Wilson loop.
The underlying relation to the supergroup may go beyond the accidental relation after the localization techniques and plays some roles in understanding the ABJ(M) theory itself.

In the supersymmetric theory, it was found that when we compute the partition function or the correlation function preserving supersymmetries, the infinite-dimensional path integral reduces to a finite-dimensional matrix integration \cite{P}.
For more careful explanation of the localization techniques, see a review \cite{Hosomichi} in this volume.

After the localization techniques, the partition function of the ABJ(M) theory is reduced to a matrix model \cite{KWY}.
Although many works are concentrated on the matrix model expression itself, solving the matrix model is an interesting and important subject.
In physics solving one theory classically means finding the exact solutions to the equation of motion, while solving one theory on the quantum-mechanical level means performing the path integral exactly.
Since the localization techniques already reduce the infinite-dimensional path integral into a matrix integration, solving the ABJ(M) theory amounts to performing the matrix integration exactly.
This is our main results in \cite{HMO1,HMO2,HMO3,HMMO} and we shall explain it in more details in the remaining part of the introduction.

Before explaining the details, let us try to discuss our motivation and expectation for the results.
First of all, the full expression of the ABJ(M) partition function may provide some hidden structures of the M2-branes.
After reproducing the predicted $N^{3/2}$ behavior in the large $N$ limit from the matrix model \cite{DMP1}, it is natural to ask how the corrections look like.
Especially, compared with the matrix interpretation of the $N^2$ behavior for the D-branes, it is perplexing to find the $N^{3/2}$ behavior for the M2-branes.
We expect that finding out all the corrections will give us some hints to this mysterious $N^{3/2}$ behavior of the M2-branes.
Prior to the discovery of the ABJ(M) theory, there were attempts in finding out the worldvolume theory of the M2-branes from a generalization of the Lie algebra \cite{BL,G}.
Though it turned out the algebra is very restricted and is not explored extensively any more, these works may suggest a hidden novel structure in M2-branes to be discovered.

Secondly, the ABJ(M) theory is one of the maximally supersymmetric theories.
If we consider that the uniqueness of string theory stems from the maximum in the supersymmetrization, we expect that the maximally supersymmetric theories will play a crucial role in understanding string theory.
At this point it should be important to study carefully all of fundamental quantities in the maximally supersymmetric theories.

Thirdly, let us stress the importance of the exact solution.
The main method in studying physics is the perturbation theory.
However, in a standard situation, the physics beyond the perturbation theory has a very rich structure.
We expect that the exact instanton expansion will provide us a window to understand non-perturbative effects.

In the remaining part of the introduction, let us shortly recapitulate the main results of a series of our works \cite{HMO1,HMO2,HMO3,HMMO}.
We will see that the results provide some insights to the above questions.

\subsection{ABJ(M) matrix model}
After applying the localization techniques, the partition function of the ABJ(M) theory reduces to a matrix model
\begin{align}
Z_k(N_1,N_2)
&=\frac{(-1)^{\frac{1}{2}N_1(N_1-1)+\frac{1}{2}N_2(N_2-1)}}{N_1!N_2!}
\nonumber\\
&\quad\int_{{\mathbb R}^{N_1+N_2}}
\prod_{i=1}^{N_1}D\mu_i\prod_{j=1}^{N_2}D\nu_j
\frac{\prod_{i<j}^{N_1}(2\sinh\frac{\mu_i-\mu_j}{2})^2
\prod_{i<j}^{N_2}(2\sinh\frac{\nu_i-\nu_j}{2})^2}
{\prod_{i,j}^{N_1,N_2}(2\cosh\frac{\mu_i-\nu_j}{2})^2},
\label{abjm}
\end{align}
where $\prod_{i<j}^{N_1}$ and $\prod_{i<j}^{N_2}$ denotes the product taking over all of pairs $(i,j)$ satisfying $1\le i<j\le N_1$ and $1\le i<j\le N_2$ respectively, while the product $\prod_{i,j}^{N_1,N_2}$ denotes $\prod_{i=1}^{N_1}\prod_{j=1}^{N_2}$.
The integration is defined by
\begin{align}
D\mu_i=\frac{d\mu_i}{2\pi}e^{\frac{ik}{4\pi}(\mu_i)^2},\quad
D\nu_j=\frac{d\nu_j}{2\pi}e^{-\frac{ik}{4\pi}(\nu_j)^2}.
\label{integral}
\end{align}
The derivation of this partition function is originally done in \cite{KWY}.
Here we shall not go into the details of the derivation and instead take this ABJ(M) matrix model \eqref{abjm} as our starting point.

\subsection{Supergroup structure}
This matrix model comes from the maximally superconformal Chern-Simons theory.
Let us shortly comment on the beauty of this matrix model \cite{DT}.
In \eqref{abjm} the most complicated part of this partition function is probably the measure
\begin{align}
\frac{\prod_{i<j}^{N_1}(2\sinh\frac{\mu_i-\mu_j}{2})^2
\prod_{i<j}^{N_2}(2\sinh\frac{\nu_i-\nu_j}{2})^2}
{\prod_{i,j}^{N_1,N_2}(2\cosh\frac{\mu_i-\nu_j}{2})^2}.
\end{align}
This measure can be simplified largely by dropping one factor of the gauge group
\begin{align}
\prod_{i<j}^{N_1}\Bigl(2\sinh\frac{\mu_i-\mu_j}{2}\Bigr)^2.
\end{align}
This measure is the famous one of the Chern-Simons matrix model.
If the reader is not familiar with it, she can replace the hyperbolic function by a rational function
\begin{align}
\prod_{i<j}^{N_1}(\mu_i-\mu_j)^2.
\end{align}
This is nothing but the $U(N_1)$ invariant Vandermonde measure appearing in the diagonalization of the $N_1\times N_1$ Hermite matrix in the Gaussian matrix model.
We can restore the hyperbolic function by exponentiating the variables $\mu$ into $e^{\pm\mu}$.
If we want to restore the dependence of the variables $\nu$, all we have to do is the supersymmetrization.
Namely we replace the invariant measure of $U(N_1)$ by that of the supergroup $U(N_1|N_2)$
\begin{align}
\frac{\prod_{i<j}^{N_1}(\mu_i-\mu_j)^2
\prod_{i<j}^{N_2}(\nu_i-\nu_j)^2}
{\prod_{i,j}^{N_1,N_2}(\mu_i+\nu_j)^2}.
\end{align}
After exponentiating $\mu_i,\nu_j$ into $e^{\pm\mu_i},e^{\pm\nu_j}$ respectively, we come back to the original measure of the ABJ(M) matrix model.
The explanation is summarized in table \ref{gaussian}.

In this sense, we can say that the ABJ(M) matrix model is the generalization of the most fundamental Gaussian matrix model with the simultaneous deformation of the supersymmetrization and the Chern-Simons deformation.
As a message, we would like to stress that, aside from the physical interpretation, even purely from a mathematical viewpoint, the ABJ(M) matrix model is one of the most fundamental matrix models which deserve careful study.

\begin{table}
\begin{eqnarray*}
&\displaystyle\frac{\prod_{i<j}^{N_1}(2\sinh\frac{\mu_i-\nu_j}{2})^2
\prod_{i<j}^{N_2}(2\sinh\frac{\nu_i-\nu_j}{2})^2}
{\prod_{i,j}^{N_1,N_2}(2\cosh\frac{\mu_i-\nu_j}{2})^2}&\\
&\nearrow\qquad\qquad\qquad\nwarrow&\\
&\qquad\qquad\qquad\displaystyle\prod_{i<j}^{N_1}\Bigl(2\sinh\frac{\mu_i-\nu_j}{2}\Bigr)^2
\qquad\qquad\qquad
\frac{\prod_{i<j}^{N_1}(\mu_i-\nu_j)^2
\prod_{i<j}^{N_2}(\nu_i-\nu_j)^2}
{\prod_{i,j}^{N_1,N_2}(\mu_i+\nu_j)^2}&\\
&\nwarrow\qquad\qquad\qquad\nearrow&\\
&\displaystyle\prod_{i<j}^{N_1}(\mu_i-\nu_j)^2&
\end{eqnarray*}
\caption{The measure of the ABJ(M) matrix model can be regarded as a simultaneous deformation of the supersymmetry deformation and the Chern-Simons deformation from the standard Vandermonde measure.}
\label{gaussian}
\end{table}

\subsection{Complete large $N$ expansion}
Now let us summarize the result found in a series of works \cite{DMP1,DMP2,FHM,O,MP,HKPT,KEK,HMO1,PY,AHS,HMO2,CM,HMO3,HMMO,H,MM,HO} for both the ABJM case and the ABJ case.
Let us reparametrize the arguments of the partition function $Z_k(N_1,N_2)$ in \eqref{abjm} as $Z_{k,M}(N)$ with $(N_1,N_2)=(N,N+M)$ and define the grand canonical partition function $\Xi_{k,M}(\mu)$ with
\begin{align}
\Xi_{k,M}(\mu)=\sum_{N=0}^\infty e^{\mu N}|Z_{k,M}(N)|,
\label{XikM}
\end{align}
and the (modified) grand potential $J_{k,M}(\mu)$ with \cite{HMO2}
\begin{align}
\Xi_{k,M}(\mu)=\sum_{n=-\infty}^\infty e^{J_{k,M}(\mu+2\pi in)},
\label{grandpot}
\end{align}
by regarding the rank $N$ as the number of particles and introducing the chemical potential $\mu$ dual to $N$.
Note that the phase dependence of the partition function $Z_{k,M}(N)$ was studied carefully in \cite{KWY} and to make contact with the topological string theory we have to take the absolute value of the partition function in defining the grand potential \cite{MM}.
It is important to notice that the large $N$ physics in the canonical ensemble is governed by the large $\mu$ limit in the grand canonical ensemble, thus below we focus on this expansion.

Note that the grand canonical partition function defined originally in \eqref{XikM} is symmetric under the $2\pi i$ shift of the chemical potential
\begin{align}
\mu\to\mu+2\pi i.
\end{align}
Due to the periodicity the grand partition function contains an oscillatory behavior in an apparent sense \cite{HMO2}.
The (modified) grand potential in \eqref{grandpot} is defined by getting rid of this oscillation.
In the inverse transformation, the partition function is given in terms of the grand canonical partition function $\Xi_{k,M}(\mu)$ by\begin{align}
|Z_{k,M}(N)|=\int_{-\pi i}^{\pi i}\frac{d\mu}{2\pi i}
\Xi_{k,M}(\mu)e^{-\mu N},
\end{align}
while in the grand potential \eqref{pertnp} by
\begin{align}
|Z_{k,M}(N)|=\int_{-\infty i}^{\infty i}\frac{d\mu}{2\pi i}
e^{J_{k,M}(\mu)-\mu N}.
\label{inverse}
\end{align}
This difference plays crucial roles, for example, in discussing the generalizations of the ABJM theory in \cite{HoMo}.

Let us first display our main result, and explain how it was found and what the physical implication is later in the subsequent subsections.
The large $\mu$ results of the grand potential can be summarized as follows.
The grand potential is split into the ``perturbative'' part and the ``non-perturbative'' part
\begin{align}
J_{k,M}(\mu)
=J_{k,M}^\text{pert}(\mu^\text{eff})
+J_{k,M}^\text{np}(\mu^\text{eff}).
\label{pertnp}
\end{align}
where ``effective'' chemical potential $\mu^\text{eff}$ will be explained later.
The perturbative part is given by
\begin{align}
J^\text{pert}_{k,M}(\mu)=\frac{C_{k}}{3}\mu^3+B_{k,M}\mu+A_{k},
\label{Jpert}
\end{align}
where the coefficients $C_{k}$ and $B_{k,M}$ are \cite{DMP1,MM}
\begin{align}
C_{k}=\frac{2}{\pi^2k},\quad 
B_{k,M}=\frac{1}{3k}-\frac{k}{12}
+\frac{k}{2}\biggl(\frac{M}{k}-\frac{1}{2}\biggr)^2,
\label{CB}
\end{align}
while $A_k$ is \cite{KEK,HaOk}
\begin{align}
A_k=\frac{2\zeta(3)}{\pi^2k}\left(1-\frac{k^3}{16}\right)
+\frac{k^2}{\pi^2}\int_0^\infty dx\frac{x}{e^{kx}-1}\log(1-e^{-2x}).
\label{Aconst}
\end{align}
For integral $k$, the integration is performed explicitly.

The non-perturbative part is given by
\begin{align}
J^\text{np}(\mu)=F^\text{top}({\bm T};g_s)
+\frac{1}{2\pi i}\frac{\partial}{\partial g_s}
\biggl[g_s F^\text{NS}\biggl(\frac{{\bm T}}{g_s};
\frac{1}{g_s}\biggr)\biggr].
\label{Jnp-to-Fref}
\end{align}
Here the functions $F^\text{top}$ and $F^\text{NS}$ are defined by the limits
\begin{align}
F^\text{top}({\bm T};\tau)&=\lim_{\tau_1\to\tau,\tau_2\to-\tau}
F^\text{ref}({\bm T};\tau_1,\tau_2),\nonumber\\
F^\text{NS}({\bm T};\tau)&=\lim_{\tau_1\to\tau,\tau_2\to 0}
2\pi i\tau_2
F^\text{ref}({\bm T};\tau_1,\tau_2),
\label{Flimit}
\end{align}
of a refined function
\begin{align}
F^\text{ref}({\bm T};\tau_1,\tau_2)
=\sum_{j_L,j_R}\sum_{n=1}^\infty\sum_{\bm d}
N^{\bm d}_{j_L,j_R}
\frac{\chi_{j_L}(q_L)\chi_{j_R}(q_R)e^{-n{\bm d}\cdot{\bm T}}}
{n(q_1^{n/2}-q_1^{-n/2})(q_2^{n/2}-q_2^{-n/2})},
\end{align}
with the su$(2)$ character
\begin{align}
\chi_j(q)=\frac{q^{2j+1}-q^{-(2j+1)}}{q-q^{-1}},
\end{align}
and various variables
\begin{align}
q_1=e^{2\pi i\tau_1},\quad
q_2=e^{2\pi i\tau_2},\quad
q_L=e^{\pi i(\tau_1-\tau_2)},\quad
q_R=e^{\pi i(\tau_1+\tau_2)}.
\end{align}
The functions in \eqref{Flimit} have their origin in topological string theory.
Namely, the function $F^\text{ref}({\bm T};\tau_1,\tau_2)$ is the free energy of the refined topological string theory, while $F^\text{top}({\bm T};\tau)$ and $F^\text{NS}({\bm T};\tau)$ are the free energy of the original topological string theory and the free energy in the so-called Nekrasov-Shatashvili (NS) limit.
The integers $N^{\bm d}_{j_L,j_R}$ are the BPS indices, which are topological invariants counting modes on the manifolds.
The relation between two K\"ahler parameters $T$ and the chemical potential $\mu$ is given by
\begin{align}
T_1=\frac{4\mu}{k}+2\pi i\biggl(\frac{1}{2}-\frac{M}{k}\biggr),\quad
T_2=\frac{4\mu}{k}-2\pi i\biggl(\frac{1}{2}-\frac{M}{k}\biggr),
\label{Kahler-T12}
\end{align}
while the string coupling $g_s$ is identified to the level $k$ as
\begin{align}
g_s=2/k.
\end{align}
Note that, in \eqref{pertnp}, we need to substitute for the effective chemical potential
\begin{align}
\mu^\text{eff}=\begin{cases}
\mu-2(-1)^{\frac{k}{2}-M}e^{-2\mu}{}_4F_3(1,1,\frac{3}{2},\frac{3}{2};
2,2,2;16(-1)^{\frac{k}{2}-M}e^{-2\mu}),&\text{for even $k$},\\
\mu+e^{-4\mu}{}_4F_3(1,1,\frac{3}{2},\frac{3}{2};2,2,2;-16e^{-4\mu}),
&\text{for odd $k$}.
\end{cases}
\label{shift}
\end{align}
It is interesting to find that the ABJ(M) matrix model has a deep relation to the topological string theory \cite{MPtop}.

This result, with the effective chemical potential plugged in, shows that the non-perturbative effects consist of terms
\begin{align}
J^\text{np}(\mu)
=\sum_{\substack{m,\ell=0\\(m,\ell)\ne(0,0)}}^\infty
f_{m,\ell}(\mu)e^{-(\frac{4m}{k}+2\ell)\mu},
\label{fml}
\end{align}
with $f_{m,\ell}(\mu)$ being polynomials of $\mu$.
Here the effect of $e^{-\frac{4\mu}{k}}$ is identified with the worldsheet instanton, which is a fundamental string wrapping the holomorphic cycle $\mathbb{CP}^1$ in ${\mathbb C}^4/{\mathbb Z}_k$ on the gravity side \cite{WSinst,DMP1}, and that of $e^{-2\mu}$ is the membrane instantons, which is a D2-brane wrapping the Lagrangian submanifold $\mathbb{RP}^3$ \cite{DMP2}.
In particular, the coefficients $f_{m,\ell}(\mu)$ with $\ell=0$ or $m=0$ take the following form \cite{MP}
\begin{align}
f_{m,0}(\mu)=d_m(k),\quad
f_{0,\ell}(\mu)=a_\ell(k) \mu^2+b_\ell(k)\mu+c_\ell(k).
\label{eq:WSM2}
\end{align}
Besides the worldsheet instantons and the membrane instantons, the full instanton series \eqref{fml} contains the bound states of these two instantons.

\subsection{History}\label{history}
Here we shall explain how the result \eqref{pertnp} with \eqref{Jpert} and \eqref{Jnp-to-Fref} was found in chronological order.
From the AdS/CFT correspondence, the $N^{3/2}$ behavior of the degrees of freedom of the $N$ multiple M2-branes \cite{KT} in the large $N$ limit was known from the gravity side about two decades ago.
The study from the gauge theory side to reproduce the $N^{3/2}$ behavior and investigate the corrections, on the other hand, was started quite recently \cite{DMP1} after the discovery of the worldvolume theory of the multiple M2-branes \cite{ABJM}.
For the explanation, we shall mainly concentrate on the ABJM case.

Since the partition function was reduced to a matrix integration from the localizations of the supersymmetric theories \cite{KWY,J,HHL}, we can apply 't Hooft expansion, the standard matrix model technique to it.
The 't Hooft expansion is to consider the 't Hooft limit
\begin{align}
N\to\infty,\quad\lambda=N/k:\text{fixed},
\label{tHooftregime}
\end{align}
and keep track of the $1/N$ corrections.
Then, the free energy $F(N,\lambda)=\log Z_{k,0}(N)$ is given by
\begin{align}
F(N,\lambda)=N^2F_0(\lambda)+F_1(\lambda)+N^{-2}F_2(\lambda)+\cdots,
\label{eq:genus}
\end{align}
which can be regarded as the genus expansion.
From this viewpoint, it is not difficult to imagine that the 't Hooft expansion has a good interpretation in string theory.

The first important work of the ABJM matrix model is the reproduction of the $N^{3/2}$ behavior \cite{DMP1} in the 't Hooft limit.
In \cite{DMP1,DMP2} it was further noted that the expansion has infinitely many $1/N$ corrections
\begin{align}
F_0(\lambda)
&=\lambda^{-\frac{1}{2}}+{\mathcal O}(e^{-\sqrt{\lambda}}),
\nonumber\\
F_1(\lambda)
&=\lambda^\frac{3}{2}+\log\lambda+{\mathcal O}(e^{-\sqrt{\lambda}}),
\nonumber\\
F_2(\lambda)
&=\lambda^\frac{7}{2}+\lambda^2+\lambda^\frac{1}{2}
+{\mathcal O}(e^{-\sqrt{\lambda}}),\label{Fexpand}\\
&\cdots,\nonumber
\end{align}
where we only present the result schematically.
Especially, the coefficients are not displayed and the 't Hooft coupling constant is slightly shifted from the original one, but we will only come back to these points later.
The study of the $1/N$ corrections is most easily done by observing the relation between the ABJ(M) matrix model and the topological string theory on local ${\mathbb P}^1\times{\mathbb P}^1$ \cite{MPtop} and studying with the holomorphic anomaly equation \cite{BCOV} from the topological string theory.
Here it is interesting to notice a triangular structure in the expansion \eqref{Fexpand}.
Namely, by going to one higher genus from $F_g(\lambda)$ to $F_{g+1}(\lambda)$ we always gain only one more perturbative term and the perturbative terms in $F_{g+1}(\lambda)/F_g(\lambda)$ range between $\lambda^2$ and $\lambda^{\frac{1}{2}}$ in the large $\lambda$ limit.
This means that the large $N$ expansion can be alternatively obtained by taking the large $N$ limit with $\lambda^2N^{-2a}$ $(0\le a\le 1)$ fixed.\footnote{See \cite{AFH} for more discussions on the limits.}
If we look at the coefficients carefully, it is not difficult to observe that after introducing \cite{DMP2}
\begin{align}
\lambda^\text{ren}=\lambda-\frac{1}{24}-\frac{\lambda^2}{3N^2},
\end{align}
the infinite expansion can be taken care of partially.
So the remaining question is whether we can perform the other summation by dropping the non-perturbative terms and solving the perturbative holomorphic anomaly equation.
In \cite{FHM} it was found that if we drop the non-perturbative corrections all of the perturbative expansions can be summed up to the Airy function
\begin{align}
Z^\text{pert}_k(N)
=e^{A_k}C_k^{-\frac{1}{3}}\Ai\Bigl[C_k^{-\frac{1}{3}}(N-B_{k,0})\Bigr].
\end{align}
In \cite{FHM} the constant $A_k$ was not taken care of but this constant was later pointed out in \cite{KEK} and studied carefully in \cite{HaOk}.
There are many arguments on the meaning of the Airy function appearing in the summation of the perturbative corrections.\footnote{See a recent work\cite{AYZ} for interesting generalizations and discussions.}
Technically it suggests the viewpoint of the grand canonical ensemble.
Namely, the integral representation of the Airy function is
\begin{align}
\Ai(N)=\int_C\frac{d\mu}{2\pi i} e^{\frac{1}{3}\mu^3-\mu N},
\end{align}
where the integral contour $C$ comes from the infinity in the region with angle $-\pi/2<\theta<-\pi/3$ seen from the origin and goes to the infinity in the region with angle $\pi/3<\theta<\pi/2$.
Comparing this expression with the inverse transformation \eqref{inverse}, it is not difficult to find that the complicated Airy function is mapped to a simple cubic function \eqref{Jpert}.
This suggests that the full expression is clearer in the grand canonical ensemble.

As pointed out in \cite{HKPT}, there are dissatisfactions with the 't Hooft expansion.
In the 't Hooft expansion, the large $N$ limit is taken by fixing the 't Hooft coupling $\lambda=N/k$.
However, in approaching to the M-theory regime in the large $N$ limit, the parameter we want to fix is the level $k$ itself which characterizes the M-theory background ${\mathbb C}^4/{\mathbb Z}_k$,
\begin{align}
N\to\infty,\quad k:\text{fixed}.
\label{Mregime}
\end{align}

These ideas were incarnated in \cite{MP}, by rewriting the partition function of the ABJM theory with gauge group $U(N)_k\times U(N)_{-k}$ into that of a Fermi gas system with $N$ non-interacting particles whose Hamiltonian is given by
\begin{align}
e^{-H}=\frac{1}{\sqrt{2\cosh\frac{q}{2}}}
\frac{1}{2\cosh\frac{p}{2}}\frac{1}{\sqrt{2\cosh\frac{q}{2}}},
\label{densitymat}
\end{align}
with the canonical commutation relation $[q,p]=i\hbar$ where the Chern-Simons level plays the role of the Planck constant, $\hbar=2\pi k$.

This beautiful rewriting directly suggests another analysis for the study of the ABJM matrix model.
Namely, we can study the grand canonical ensemble of the ABJM matrix model from the WKB expansion around $\hbar=0$ \cite{MP}.
In this expansion, we first expand the whole grand potential in the small $k$ region
\begin{align}
J_k(\mu)=\frac{1}{k}J^{(0)}(\mu)+kJ^{(2)}(\mu)+k^3J^{(4)}(\mu)+\cdots,
\label{Jexp}
\end{align}
and further study each term separately in the large $\mu$ expansion.
After that, we collect all terms of the same order in $\mu$ and hope that we can sum up the $k$ expansion.
In this sense we can separate the small $k$ expansion and the large $\mu$ expansion and take the large $\mu$ limit by keeping the level $k$ fixed. 

Both the 't Hooft expansion and the WKB expansion are powerful to study the matrix model to some extent.
On one hand, the 't Hooft expansion detects not only the perturbative part of the Airy function but also the worldsheet instanton part, though is sensible to neither the membrane instanton part nor the bound state part.
This is because after plugging in the stationary condition for $\mu$
\begin{align}
N=\frac{\partial J_k}{\partial\mu}\simeq C_k\mu^2,
\label{stationary}
\end{align}
the worldsheet instanton factor and the membrane instanton factor in \eqref{fml} become
\begin{align}
e^{-\frac{4\mu}{k}}\simeq e^{-2\pi\sqrt{2\lambda}},\quad
e^{-2\mu}\simeq e^{-\pi N\sqrt{2/\lambda}}.
\end{align}
On the other hand, the WKB expansion detect only the perturbative part and the membrane instanton part, but not the worldsheet instanton part or  the bound state part.
So, unfortunately, for the bound state of the worldsheet instantons and the membrane instantons, neither the 't Hooft expansion nor the WKB expansion is applicable.
The only method we have is the numerical analysis and the pole cancellation mechanism, which we will explain in the next subsection.

\subsection{Pole cancellation mechanism}\label{cancellation}

In determining the non-perturbative effects, the so-called pole cancellation mechanism plays an important role.
Let us recapitulate it shortly.
For this purpose we display the explicit forms of the first few worldsheet instantons in \eqref{eq:WSM2} 
\begin{align}
d_1(k)=\frac{1}{\sin^2\frac{2\pi}{k}},\quad
d_2(k)=-\frac{1}{2\sin^2\frac{4\pi}{k}}-\frac{1}{\sin^2\frac{2\pi}{k}},
\quad\cdots,
\label{afewWS}
\end{align}
and membrane instantons \cite{HMO2,CM}
\begin{align}
a_1(k)&=-\frac{4\cos\frac{\pi k}{2}}{\pi^2k},&
b_1(k)&=\frac{2\cos^2\frac{\pi k}{2}}{\pi\sin\frac{\pi k}{2}},&
c_1(k)&=\cdots,\nonumber\\
a_2(k)&=-\frac{8+10\cos\pi k}{\pi^2 k},&
b_2(k)&=\frac{4(1+\cos\pi k)}{\pi^2 k}
+\frac{17+24\cos\pi k}{2\pi\sin\pi k},&
c_2(k)&=\cdots,\nonumber\\
&&&\cdots.&&
\label{afewMB}
\end{align}
It is not difficult to find that the coefficients of the $m$-th worldsheet instanton contain poles when $\frac{2m}{k}\in{\mathbb Z}$, while the coefficients of the $\ell$-th membrane instanton contain poles when $2\ell k\in{\mathbb Z}$.
However, it is interesting to observe that whenever one coefficient is divergent the other coefficient is always divergent as well and totally the coefficient is completely finite.
This property was first observed in \cite{HMO2} and the importance was stressed in \cite{CM}.

This property is probably due to the supersymmetries.
Since there is no phase transition in theories with enough supersymmetries, the partition function should be finite.
From the viewpoint of applications, this property can be used to fix first few instanton coefficients.
In fact the first few coefficients of the membrane instantons listed above were found from this pole cancellation mechanism.
The existence of the membrane instantons was first observed in \cite{BBS}.
It is surprising that the first quantitative study of the membrane instanton is performed in this way.

We can proceed to the bound states and find a similar pole cancellation mechanism there.
Namely, instead of the simple pole cancellation mechanism between the worldsheet instanton and the membrane instanton, for higher instanton effects the bound state of the worldsheet instanton and the membrane instanton appears as well.
Hence, we need to consider the cancellation among all of the bound states besides the pure worldsheet instanton and the pure membrane instanton.

In \cite{HMO3} we find that the effect of the bound states can be incorporated into that of the worldsheet instantons by shifting the chemical potential as in \eqref{shift}.
As a bonus, after shifting the chemical potential $\mu$ into the effective one $\mu_\text{eff}$, we find that the quadratic polynomial coefficients of the pure membrane instanton $f_{0,\ell}(\mu)$ \eqref{eq:WSM2} become linear polynomials, where the coefficients of the linear terms and constant terms obey a derivative relation.
Finally, in \cite{HMMO}, all of these structure are reexpressed in terms of the refined topological string theory as in \eqref{pertnp}.

\subsection{Contents}

The main purpose of this paper is to review the beautiful aspects of the ABJ(M) matrix model and its generalizations.
After reviewing the history of the progress in section \ref{history} and section \ref{cancellation}, it is not difficult to imagine that the explanation would be the simplest if we start with reviewing the Fermi gas formalism.
We shall do this in the next section first for the ABJM matrix model and then generalize to the ABJ matrix model.
After that, in section \ref{npcorr}, we shall proceed to various analysis including the WKB expansion, the exact evaluation of the partition function, the numerical study of the grand potential, the pole cancellation mechanism and the viewpoint from the topological string theory.
Finally, in section \ref{topics} we summarize some recent progress in the supersymmetry enhanced case and more general cases.

\section{Fermi gas formalism}
In this section, let us explain how the Fermi gas formalism is applied to the ABJM and ABJ matrix model.

\subsection{ABJM matrix model}
In this subsection, we shall explain the Fermi gas formalism for the ABJM matrix model \cite{MP}.
Namely, we shall study \eqref{abjm} by restricting ourselves to the case of $N_2=N_1=N$ and define
\begin{align}
Z_k(N)=Z_k(N,N).
\end{align}
Using the Cauchy identity
\begin{align}
\det\left(\frac{1}{u_i+v_j}\right)
=\frac{\prod_{i<j}(u_i-u_j)(v_i-v_j)}{\prod_{i,j}(u_i+v_j)},
\label{Cauchy}
\end{align}
one can show that \eqref{abjm} is rewritten as
\begin{align}
Z_k(N)=\int\frac{D^N\mu}{N!}\frac{D^N\nu}{N!}
\det\biggl(\frac{1}{2\cosh\frac{\mu_i-\nu_j}{2}}\biggr)
_{\begin{subarray}{c}1\le i\le N\\1\le j\le N\end{subarray}}
\det\biggl(\frac{1}{2\cosh\frac{\nu_i-\mu_j}{2}}\biggr)
_{\begin{subarray}{c}1\le i\le N\\1\le j\le N\end{subarray}}.
\end{align}
Note that the two determinant factors are identical, though we write them separately, so that the structure is clearer.
Using a determinantal formula in \cite{MM}, we can combine the two determinants
\begin{align}
Z_k(N)=\int\frac{D^N\mu}{N!}\det
\biggl(\frac{1}{2\cosh\frac{\mu_i-\nu}{2}}\circ
\frac{1}{2\cosh\frac{\nu-\mu_j}{2}}\biggr)
_{\begin{subarray}{c}1\le i\le N\\1\le j\le N\end{subarray}}.
\label{Zdet}
\end{align}
Note that here $\circ$ stands for the integration
\begin{align}
\phi\circ\psi=\int D\nu\phi(\nu)\psi(\nu).
\label{contract}
\end{align}
In other words, we regard two ingredients $\bigl(2\cosh\frac{\mu-\nu}{2}\bigr)^{-1}$ and $\bigl(2\cosh\frac{\nu-\mu}{2}\bigr)^{-1}$ as matrices with the continuous indices $\mu$ and $\nu$.
As in the discrete case, where we perform the matrix multiplication by contracting the indices, here the contractions of the continuous indices are done by the integrations given in \eqref{integral}.

After expanding the determinant of \eqref{Zdet} and rescaling the integration variables $\mu=x/k$, $\nu=y/k$, we find that it can be expressed as a sum over permutations $S_N$
\begin{align}
Z_k(N)=\frac{1}{N!}\sum_{\sigma\in S_N}(-1)^\sigma
\int\frac{d^Nx}{(2\pi)^N}
\prod_{i=1}^N\rho(x_i,x_{\sigma(i)}),
\label{tosigma}
\end{align}
with $\rho(x,x')$ given by
\begin{align}
\rho(x,x')=\int\frac{dy}{2\pi}e^{\frac{i}{4\pi k}x^2}
\frac{1}{k}\frac{1}{2\cosh\frac{x-y}{2k}}
e^{-\frac{i}{4\pi k}y^2}
\frac{1}{k}\frac{1}{2\cosh\frac{y-x'}{2k}}.
\end{align}
If we use the Fourier transformation
\begin{align}
\frac{1}{k}\frac{1}{2\cosh\frac{x-y}{2k}}
=\langle x|\frac{1}{2\cosh\frac{\widehat p}{2}}|y\rangle,
\end{align}
by introducing the coordinate operator $\widehat q$ and the momentum operator $\widehat p$ satisfying $[\widehat q,\widehat p]=i\hbar$ with $\hbar=2\pi k$ and the coordinate eigenstate 
\begin{align}
\widehat q|x\rangle=x|x\rangle,
\end{align}
normalized by $\langle x|x'\rangle=2\pi\delta(x-x')$, we find that $\rho(x,x')$ can be expressed as
\begin{align}
\rho(x,x')=\langle x|e^{\frac{i}{2\hbar}\widehat q^2}
\frac{1}{2\cosh\frac{\widehat p}{2}}
e^{-\frac{i}{2\hbar}\widehat q^2}
\frac{1}{2\cosh\frac{\widehat p}{2}}|x'\rangle.
\end{align}
After some further canonical transformations with the help of the formula
\begin{align}
e^{\frac{i}{2\hbar}\widehat q^2}
f(\widehat p)
e^{-\frac{i}{2\hbar}\widehat q^2}
=f(\widehat p-\widehat q),\quad
e^{\frac{i}{2\hbar}\widehat p^2}
g(\widehat q)
e^{-\frac{i}{2\hbar}\widehat p^2}
=g(\widehat q+\widehat p),
\end{align}
and a similarity transformation, we finally find that $\rho(x,y)$ can be interpreted as a density matrix of a quantum mechanical system
\begin{align}
\rho(x,y)=\langle x|e^{-H}|y\rangle,
\end{align}
with \eqref{densitymat}, which leads to the expression
\begin{align}
\rho(x,y)=\frac{1}{k}
\frac{1}{\sqrt{2\cosh\frac{x}{2}}}\frac{1}{2\cosh\frac{x-y}{2k}}
\frac{1}{\sqrt{2\cosh\frac{y}{2}}}.
\label{density}
\end{align}
The formula \eqref{tosigma} can be regarded as the $(-1)^\sigma$-graded partition function of a particle propagating from a state labeled by the coordinate $x_i$ to the permutated coordinate $x_{\sigma(i)}$.
This is nothing but the partition function of a Fermi gas system.
To avoid unnecessary complexities, hereafter we shall often drop the hats for the operators $\widehat q$ and $\widehat p$.

It is convenient to define the grand partition function $\Xi_k(\mu)$ by introducing the chemical potential $\mu$ conjugate to $N$ and summing over $N$.
It turns out that $\Xi_k(\mu)$ can be written as a Fredholm determinant of $\rho$
\begin{align}
\Xi_k(\mu)=\sum_{N=0}^\infty e^{N\mu}Z_k(N)=\det(1+e^\mu\rho),
\label{Fredholm}
\end{align}
and the grand potential $\widetilde J_k(\mu)=\log\Xi_k(\mu)$ is given by\footnote{Note that this grand potential $\widetilde J_k(\mu)$ is different from the modified one defined in \eqref{grandpot}.}
\begin{align}
\widetilde J_k(\mu)=\sum_{\ell=1}^\infty\frac{(-1)^{\ell-1}}{\ell}e^{\ell\mu}\Tr\rho^\ell.
\label{Jtrrho}
\end{align}
The expression of the Fredholm determinant \eqref{Fredholm} and its expansion \eqref{Jtrrho} can be shown by relating the coefficients of $\prod_{\ell=1}^\infty(\Tr\rho^\ell)^{m_\ell}$, to the combinatorics of distributing $N$ into an assembly of $m_\ell$ cycles of length $\ell$ satisfying $\sum_{\ell=1}^\infty m_\ell\ell=N$ \cite{statistical}.

In terms of the eigenvalues $E_n~(n=0,1,\cdots)$ of the Hamiltonian $H$, the grand partition function is given by
\begin{align}
\Xi_k(\mu)=\prod_{n=0}^\infty(1+e^{\mu-E_n}),
\label{QM}
\end{align}
hence the problem boils down to the diagonalization of the Hamiltonian.
The study of the spectrum of the ABJM Hamiltonian was initiated in \cite{HMO1} using the Hankel matrix representation of $\rho$, which was later generalized to other models.

Note also that the partition function can also be rewritten as
\begin{align}
Z_k(N)=\frac{1}{N!}\int\frac{d^Nx}{(4\pi k)^N}
\prod_{i<j}^N\tanh^2\frac{x_i-x_j}{2k}
\prod_{i=1}^N\frac{1}{2\cosh\frac{x_i}{2}},
\label{ZABJM}
\end{align}
which is useful to discuss the generalization to the ABJ model.

\subsection{ABJ matrix model}
In the previous subsection, we have given the Fermi gas formalism for the ABJM matrix model.
Using this Fermi gas formalism, we shall explain how the exact instanton expansion was found in the next section.
Before it, let us comment that the Fermi gas formalism is also applicable to the ABJ matrix model where the gauge group $U(N_1)_k\times U(N_2)_{-k}$ has different ranks.
In generalizing the discussions in the previous subsection, there are two Fermi gas formalisms for the ABJ matrix model.
One is to dress the density matrix in \eqref{density} by extra factors with the traces kept unchanged, while the other is to keep the density matrix and introduce the endpoints for the density matrix.

Since the extra physical ingredients of the ABJ theory compared with the ABJM theory is the fractional M2-branes, two formalisms amount to expressing the fractional branes by changing the closed string backgrounds or from the open string excitations.
Hence, it is reasonable to call these two formalisms as closed string formalism and open string formalism respectively.

After the perturbative coefficients $C_k$, $B_{k,M}$ \eqref{CB} and $A_k$ \eqref{Aconst} and the worldsheet instantons were found from the consistency with the 't Hooft expansion \cite{DMP1}, the topological string on local $\mathbb{P}^1\times\mathbb{P}^1$ and the ABJM limit $M\to0$, we can use either of these Fermi gas formalisms to further study the membrane instantons and the bound states.

\subsubsection{Closed string formalism}
Let us start the study of the ABJ matrix model \eqref{abjm} with different ranks.
In what follows, we will set $N_1=N, N_2=N+M$ and assume $k,M>0$ without loss of generality.
First let us note that the partition function for $N=0$ is nothing but the bosonic $U(M)$ Chern-Simons matrix model, which is given by \cite{jones}
\begin{align}
Z_k(0,M)=k^{-\frac{M}{2}}
\prod_{s=1}^{M-1}\left(2\sin\frac{\pi s}{k}\right)^{M-s}.
\end{align}
As discussed in \cite{AHS,H,HO}, if we define $\widehat Z_k(N,N+M)$ to be the absolute value of the partition function $Z_k(N,N+M)$ normalized by the Chern-Simons case ($N=0$),
\begin{align}
\widehat Z_k(N,N+M)=\biggl|\frac{Z_k(N,N+M)}{Z_k(0,M)}\biggr|,
\end{align}
we find that $\widehat Z_k(N,N+M)$ is given by an integration
\begin{align}
\widehat{Z}_k(N,N+M)&=\frac{1}{N!}\int\frac{d^Nx}{(4\pi k)^N}
\prod_{i<j}^N\tanh^2\frac{x_i-x_j}{2k}\prod_{i=1}^NV(x_i),
\end{align}
with a dressing factor
\begin{align}
V(x)&=\frac{1}{e^{\frac{x}{2}}+(-1)^Me^{-\frac{x}{2}}}\prod_{s=-\frac{M-1}{2}}^{\frac{M-1}{2}}
\tanh\frac{x+2\pi is}{2k},
\end{align}
which is very similar to \eqref{ZABJM} in the ABJM matrix model.
The Seiberg-like duality of the ABJ theory
\begin{align}
U(N)_k\times U(N+M)_{-k}~\Leftrightarrow~ U(N)_{-k}\times U(N+k-M)_{k}
\end{align}
implies that $\widehat{Z}_k(N,N+M)$ is invariant under the exchange $M\leftrightarrow k-M$ with $k$ and $N$ fixed.
Also, note that the original partition function $Z_k(N,N+M)$ (as well as the Chern-Simons case $Z_k(0,M)$) vanishes when $k<M$, which is consistent with the conjecture that the supersymmetry of the ABJ theory is spontaneously broken for $k<M$. 

Again, using the Cauchy identity, $\widehat{Z}_k(N,N+M)$ can be recast as a Fermi gas form
\begin{align}
\widehat{Z}_k(N,N+M)&=\frac{1}{N!}\sum_{\sigma\in S_N}(-1)^\sigma
\int\frac{d^Nx}{(2\pi)^N}\prod_{i=1}^N\rho(x_i,x_{\sigma(i)}),
\end{align}
with the density matrix \eqref{density} changed into
\begin{align}
\rho(x,y)&=\frac{1}{k}\sqrt{V(x)}\frac{1}{2\cosh\frac{x-y}{2k}}\sqrt{V(y)}.
\label{rhoABJ}
\end{align}
Using this dressed density matrix, the grand partition function $\Xi_{k,M}(\mu)$ in \eqref{XikM} is given by
\begin{align}
\Xi_{k,M}(\mu)
=|Z_{k}(0,M)|\det(1+e^\mu\rho).
\label{XiABJ}
\end{align}

\subsubsection{Open string formalism}
Having seen the description of the fractional branes by changing the closed string background in the previous subsubsection, here let us turn to the description from the open string \cite{MM}.
This formalism has a benefit in the compatibility with the vacuum expectation values of the half-BPS Wilson loop, since we end up with an ultimate Fermi gas formalism which includes both the M2 fractional branes and the half-BPS Wilson loop.
As a corollary of this formalism, we can reproduce, on one hand, the Fermi gas formalism for the vacuum expectation values of the half-BPS Wilson loop \cite{HHMO}, and on the other hand, that for the partition function of the M2 fractional branes \cite{MM}.

To state the results we need some preparations.
Instead of the partition function itself, since the formalism incorporates the insertion of the half-BPS Wilson loop as well, let us consider the vacuum expectation value of the half-BPS Wilson loop.
After the localization of the supersymmetric theory, it was found \cite{KWY,DT} that the insertion means to include a loop operator in the matrix model
\begin{align}
\langle s_Y\rangle_{k}(N_1,N_2)
&=\frac{(-1)^{\frac{1}{2}N_1(N_1-1)\frac{1}{2}N_2(N_2-1)}}{N_1!N_2!}
\int\frac{d^{N_1}\mu}{(2\pi)^{N_1}}\frac{d^{N_2}\nu}{(2\pi)^{N_2}}
s_Y(e^{\mu}|e^{\nu})
\nonumber\\
&\times\left[\frac{\prod_{i<j}^{N_1}2\sinh\frac{\mu_i-\mu_j}{2}
\prod_{i<j}^{N_2}2\sinh\frac{\nu_i-\nu_j}{2}}
{\prod_{i=1}^{N_1}\prod_{j=1}^{N_2}2\cosh\frac{\mu_i-\nu_j}{2}}\right]^2
e^{\frac{ik}{4\pi}(\sum_{i=1}^{N_1}\mu_i^2-\sum_{j=1}^{N_2}\nu_j^2)}.
\end{align}
Here the loop operator 
$s_Y(e^{\mu}|e^{\nu})=s_Y(e^{\mu_1},e^{\mu_2},\cdots,e^{\mu_{N_1}}|
e^{\nu_1},e^{\nu_2},\cdots,e^{\nu_{N_2}})$ is the so-called supersymmetric Schur polynomial, which is nothing but the character of the hidden supergroup $U(N_1|N_2)$.
This is a symmetric polynomial with $N_1$ variables $\mu_i$ and $N_2$ variables $\nu_j$, labeled by a Young tableaux $Y$.
It is probably the simplest to find out this supersymmetric Schur polynomial by expressing the original Schur polynomial in terms of a diagonal matrix $\diag(e^{\mu_i})_{i=1}^{N_1}$ and the trace $\tr$ (in other words, the power sum symmetric polynomials) and replacing the diagonal matrix by a diagonal supermatrix $\diag((e^{\mu_i})_{i=1}^{N_1}|(-e^{\nu_j}){}_{j=1}^{N_2})$ and the trace by a supertrace.
For example, let us consider the supersymmetric Schur polynomial of the second symmetric tensor in the supergroup $U(N_1|N_2)$ with $N_1$ and $N_2$ being large enough
\begin{align}
s_{\Young[-1]{2}}(e^\mu)=\frac{1}{2}(\tr\diag(e^{\mu}))^2
+\frac{1}{2}\tr\diag(e^{2\mu}).
\end{align}
Then after the replacement we find that the supersymmetric Schur polynomial can be decomposed into the original one by
\begin{align}
s_{\Young[-1]{2}}(e^\mu|e^\nu)=s_{\Young[-1]{2}}(e^\mu)
+s_{\Young[-1]{1}}(e^\mu)s_{\Young[-1]{1}}(e^\nu)
+s_{\Young[-2]{11}}(e^\nu).
\end{align}

Also, let us define the vacuum expectation value in the grand canonical ensemble
\begin{align}
[s_Y]^\text{GC}_{k,M}(z)=\sum_{N=0}^\infty\langle s_Y\rangle_k(N,N+M)z^N,
\end{align}
as for the partition function.
We further normalize it by the ABJM grand canonical partition function
\begin{align}
\langle s_Y\rangle^\text{GC}_{k,M}(z)
=\frac{[s_Y]^\text{GC}_{k,M}(z)}{[1]^\text{GC}_{k,0}(z)}.
\label{normGC}
\end{align}
Then, we can prove
\begin{align}
\langle s_Y\rangle^\text{GC}_{k,M}(z)=\det\bigl(H_{p,q}(z)\bigr)
_{\begin{subarray}{c}1\le p\le M+r\\1\le q\le M+r\end{subarray}},
\label{giambellivev}
\end{align}
where $H_{p,q}(z)$ is given by
\begin{align}
H_{p,q}(z)=\begin{cases}
E_{l_p}(1+zQP)^{-1}E_{-M+q-1},&\text{for}\;1\le q\le M,\\
zE_{l_p}(1+zQP)^{-1}QE_{a_{q-M}},&\text{for}\;1\le q-M\le r.
\end{cases}
\label{Hpq}
\end{align}

We have to explain the meaning of each expression.
First of all, we regard $(Q)_{\nu,\mu}$ and $(P)_{\mu,\nu}$ as matrices and $(E_j)_\nu$ as a vector with continuous indices $\mu$ and $\nu$.
Again, as explained below \eqref{contract}, we multiply matrices with continuous indices by the integrations \eqref{integral}.
So each expression is a scalar constructed by the inner product of two vectors with many matrices sandwiched in between.
The explicit form of $(Q)_{\nu,\mu}$, $(P)_{\mu,\nu}$ and $(E_j)_\nu$ is
\begin{align}
(Q)_{\nu,\mu}=\frac{1}{2\cosh\frac{\nu-\mu}{2}},\quad
(P)_{\mu,\nu}=\frac{1}{2\cosh\frac{\mu-\nu}{2}},\quad
(E_j)_\nu=e^{(j+\frac{1}{2})\nu}.
\end{align}
Now let us turn to the integers $l_p$ and $a_q$.
In short, these integers are $p$-th leg length and $q$-th arm length, which appear in the Frobenius symbol $(a_1a_2\cdots a_r;l_1l_2\cdots l_{r+M})$.
The Frobenius symbol for the ABJM $U(N|N)$ case is found by drawing the diagonal line for the Young diagram and counting the number of boxes horizontally or vertically for the arm length or the leg length, respectively.
For the general ABJ case with $U(N|N+M)$, all we have to do is to shift the diagonal line by $M$ boxes.
Note that, after shifting the diagonal line, the number of the arm lengths and that of the leg lengths are not equal any more.
Hence, we cannot consider the determinant of the square matrix.
We can restore, however, the equality of the numbers by consider the ``inverse'' arm length as well.
The subscript $-M+q-1$ appearing in \eqref{Hpq} is nothing but this inverse arm length.
Therefore, we can say that the lower expression in \eqref{Hpq} comes from the usual arm length while the upper one is from the inverse arm length.
(See e.g.\ \cite{HHMO,MM} for figures explaining the Frobenuis symbol.) 

Before explaining how the expression is obtained, let us first see how this formula is beautiful and useful simultaneously.
First let us consider the ABJM case $N_2=N_1$ with a non-trivial insertion of the loop operator.
Then, we find that $M=0$ and the upper expression of $H_{p,q}$ is absent.
If we further use the same formula for the hook Young tableaux
\begin{align}
\langle s_{(a_q;l_p)}\rangle^\text{GC}_{k,0}
=zE_{l_p}(1+zQP)^{-1}QE_{a_{q}},
\end{align}
then, as a corollary of the above formula, we can express the vacuum expectation value of a general loop operator as the determinant of those of the hook loop operators constructed by taking all combinations of the arm lengths and the leg lengths from the original Young tableaux \cite{HHMO}
\begin{align}
\langle s_Y\rangle^\text{GC}_{k,0}(z)
=\det\bigl(\langle s_{(a_q;l_p)}\rangle^\text{GC}_{k,0}(z)\bigr)
_{\begin{subarray}{c}1\le p\le r\\1\le q\le r\end{subarray}}.
\end{align}
Note that, without taking the vacuum expectation value, it is a classical mathematical result stating that the character itself satisfies the same type of formula
\begin{align}
s_Y=\det\bigl(s_{(a_q;l_p)}\bigr)
_{\begin{subarray}{c}1\le p\le r\\1\le q\le r\end{subarray}},
\label{giambelli}
\end{align}
which is called Giambelli formula.
Hence, in other words, what we have shown is that we can put to each loop operator in the Giambelli formula \eqref{giambelli} the normalized vacuum expectation values in the sense of the grand canonical ensemble.
Namely, the normalized vacuum expectation values defined in \eqref{normGC} is Giambelli compatible.

Next, let us turn to another special case of the partition function with no insertions $s_Y=1$.
Then, we find
\begin{align}
\langle 1\rangle^\text{GC}_{k,M}(z)
=\det\bigl(E_{M-p}(1+zQP)^{-1}E_{-M+q-1}\bigr)_
{\begin{subarray}{c}1\le p\le M\\1\le q\le M\end{subarray}},
\end{align}
since $l_p=M-p$ in this case.
If we regard the trace operator without endpoints as closed strings and the meson operators $EQPQP\cdots E$ which appears by expanding the expression in \eqref{Hpq} as open strings, this formula seems to state that the fractional M2-branes can be constructed from the open strings.
This is reminiscent of the fact that classical D-brane solutions can be studied from open string field theory \cite{SZ}.

So far we have stressed the beauty of this formula.
It is also interesting to note that this formula is actually very useful.
In section \ref{exactvalues}, we shall see that the special form of the density matrix is very useful to compute the powers of the density matrix.
Our formula in \eqref{Hpq} also shows the same computability.
Namely, due to the expression of the meson operator, we can start with a vector $E$ and multiply matrices $Q$ and $P$ to it one after another without difficulty.

Let us now turn to the explanation how this expression is obtained.
First let us explain the formula for the partition function $\langle 1\rangle^\text{GC}_{k,M}(z)$.
Again our starting point is the matrix model \eqref{abjm}.
Instead of the Cauchy determinant formula \eqref{Cauchy}, let us consider the following formula
\begin{align}
\det\begin{pmatrix}\left(\displaystyle\frac{1}{u_i+v_j}\right)_{\begin{subarray}{c}
1\le i\le N_1\\1\le j\le N_2
\end{subarray}}\\
\left(v_j^{N_2-N_1-i}\right)_{\begin{subarray}{c}
1\le i\le N_2-N_1\\1\le j\le N_2
\end{subarray}}
\end{pmatrix}
=\frac{\prod_{i<j}^{N_1}(u_i-u_j)\prod_{i<j}^{N_2}(v_i-v_j)}
{\prod_{i,j}^{N_1,N_2}(u_i+v_j)}.
\label{VdmC}
\end{align}
This is a natural combination of the Vandermonde determinant and the Cauchy determinant, which reduces to the Vandermonde determinant for the special case of $N_1=0$, while reduces to the Cauchy determinant \eqref{Cauchy} for $N_2=N_1$.
This formula can be proved by starting with the Cauchy determinant \eqref{Cauchy} with matrix size $N_2\times N_2$ and sending $v_{N_2}$, $v_{N_2-1}$, $\cdots$, $v_{N_1+1}$ to infinity one after another.

\begin{table}
\begin{align*}
\det\begin{pmatrix}
\frac{1}{u_1+v_1}&
\fbox{$\frac{1}{u_1+v_2}$}&\cdots&
\frac{1}{u_1+v_{N_2}}\\
\vdots&\vdots&&\vdots\\
\frac{1}{u_{N_1}+v_1}&
\frac{1}{u_{N_1}+v_2}&\cdots&
\frac{1}{u_{N_1}+v_{N_2}}\\
\fbox{$v_1^{N_2-N_1-1}$}&
v_2^{N_2-N_1-1}&\cdots&
v_{N_2}^{N_2-N_1-1}\\
v_1^{N_2-N_1-2}&
v_2^{N_2-N_1-2}&\cdots&
v_{N_2}^{N_2-N_1-2}\\
\vdots&\vdots&&\vdots\\
v_1&v_2&\cdots&v_{N_2}\\
1&1&\cdots&1
\end{pmatrix}
\det\begin{pmatrix}
\fbox{$\frac{1}{v_1+u_1}$}&
\frac{1}{v_2+u_1}&\cdots&
\frac{1}{v_{N_2}+u_1}\\
\vdots&\vdots&&\vdots\\
\frac{1}{v_1+u_{N_1}}&
\frac{1}{v_2+u_{N_1}}&\cdots&
\frac{1}{v_{N_2}+u_{N_1}}\\
v_1^{N_2-N_1-1}&
v_2^{N_2-N_1-1}&\cdots&
v_{N_2}^{N_2-N_1-1}\\
v_1^{N_2-N_1-2}&
\fbox{$v_2^{N_2-N_1-2}$}&\cdots&
v_{N_2}^{N_2-N_1-2}\\
\vdots&\vdots&&\vdots\\
v_1&v_2&\cdots&v_{N_2}\\
1&1&\cdots&1
\end{pmatrix}
\end{align*}
\caption{An explanation how the $EQPE$ term in the expansion of $E(1+zQP)^{-1}E$ in \eqref{Hpq} appears.}
\label{twodet}
\end{table}

Then, after plugging in $u_i=e^{x_i}$ and $v_j=e^{y_j}$, we can reproduce the measure of the ABJ matrix model.
Let us see how we can put the measure into a Fredholm determinantal expression as in the case of the ABJM matrix model.
Since the substitution can be done at any stage, for simplicity, let us stick to the variables $u_i$ and $v_j$. 

Since in the partition function \eqref{abjm} we have two factors of this measure, we have two determinants.
(See table \ref{twodet}.)
So there are a lot of terms in expanding the determinant.
Suppose we pick up one element from the lower $(N_2-N_1)\times N_2$ submatrix of the first determinant, say $v_1^{N_2-N_1-1}$.
This means that, in this term, we are not allowed to pick up the other elements in the same row or in the same column.
This element $v_1^{N_2-N_1-1}$ is associated to the integration of the  variable $v_1$.
To finish the integration of $v_1$, we need to decide which element to pick up from the first column in the second determinant.

If we pick up one element from the lower $(N_2-N_1)\times N_2$ submatrix of the second determinant, say $v_1^{N_2-N_1-2}$, again this means that we do not pick up the others in the same row or column.
Then, we do not have other elements depending on the variable $v_1$ and we can complete the $v_1$ integration.

Instead, suppose in the second determinant we pick up another element from the upper $N_1\times N_2$ submatrix, say $(v_1+u_1)^{-1}$.
Then, again there are no other elements associated with the variable $v_1$ and we can integrate $v_1$.
But now we are associated with the integration of another variable $u_1$.
To finish the $u_1$ integration, we need to pick up one element from the first row of the first determinant, say $(u_1+v_2)^{-1}$.
(Note that we can no more pick up the element $(u_1+v_1)^{-1}$ from the first determinant, because we already pick up $v_1^{N_2-N_1-1}$.)
Then, we are associated with the integration of the other variable $v_2$.

Now, in the second determinant, if we choose from the lower $(N_2-N_1)\times N_2$ submatrix, then we finish the $v_2$ integration.
But if we choose from the upper $N_1\times N_2$ submatrix, we need to finish another integration of the variable $u$.

We can continue this process and find a term consisting of a series of matrix multiplications, with two monomials of $v$ on the two ends connected by products of $(v+u)^{-1}(u+v)^{-1}$.
After plugging in $u_i=e^{x_i}$ and $v_j=e^{y_j}$, essentially $v$ on the two ends turns into $E$ while $(v+u)^{-1}(u+v)^{-1}$ becomes $QP$.
This can be regarded as an open string with Chan-Paton factors on the two ends.
Of course, there are also contributions from the pure traces without ends.
Now the only problem is to count the combinatorial factor correctly.
The answer can be found more efficiently by preparing a mathematical formula for the matrix multiplications.
This was done in \cite{MM}.

For the vacuum expectation values of the loop operator, we can use another beautiful formula \cite{VdJM}, stating that the supersymmetric Schur polynomial can be expressed as a ratio of two determinants, where the determinant in the denominator is nothing but the Vandermonde-Cauchy determinant we encounter in \eqref{VdmC}.
Hence, by combining the supersymmetric Schur polynomial with the measure, we find that one of the two determinants appearing in the measure of the partition function is replaced by another determinant.
Hence, we can repeat the process explained in table \ref{twodet} and find out the formula \eqref{giambellivev}.

\section{Non-perturbative corrections at large $N$}\label{npcorr}
After presenting the Fermi gas formalism for the ABJ(M) matrix model in the previous section, we can now start our study of the large $\mu$ expansion for the grand potential.
For simplicity, we will mainly focus on the ABJM case.
Since the expression as the partition function of a Fermi gas system already suggests a statistical mechanical method, we first start with the WKB $\hbar$ expansion.
After that we turn to the study of the exact values of the partition function at finite $N$ and see how these exact values lead to the numerical study of the grand partition function.
Finally, we compare with the free energy of the topological string theory and see how the exact expression is found.

\subsection{WKB expansion}\label{wkb}
As emphasized in \cite{MP,GM}, the Fermi gas formalism enables us to study the M-theory regime \eqref{Mregime}.
One useful way to study this regime is to perform the semi-classical (or WKB) expansion \eqref{Jexp} of the grand potential $\widetilde J_k(\mu)=\log\Xi_k(\mu)$ around $k=0$ in the limit $\mu\to\infty$.

For example, the leading order term $\widetilde J^{(0)}(\mu)$ in the WKB expansion is easily found by replacing $\Tr\rho^\ell$ in \eqref{Jtrrho} by the classical phase space integral \cite{MP}
\begin{align}
\widetilde J^{(0)}(\mu)&=\sum_{\ell=1}^\infty \frac{(-1)^{\ell-1}}{\ell}e^{\ell\mu}
\int\frac{dqdp}{2\pi}
\left(\frac{1}{2\cosh\frac{q}{2}}\frac{1}{2\cosh\frac{p}{2}}\right)^\ell
\nonumber\\
&=\frac{e^\mu}{4}{}_3F_2
\left(\frac{1}{2},\frac{1}{2},\frac{1}{2};
1,\frac{3}{2};\frac{e^{2\mu}}{16}\right)
-\frac{e^{2\mu}}{8\pi^2}{}_4F_3\left(1,1,1,1;
\frac{3}{2},\frac{3}{2},2;\frac{e^{2\mu}}{16}\right).
\end{align}
From the large $\mu$ behavior of the hypergeometric functions in the above expression,
\begin{align}
&\widetilde J^{(0)}(\mu)
=\frac{2\mu^3}{3\pi^2}+\frac{\mu}{3}+\frac{2\zeta(3)}{\pi^2}\nonumber\\
&\quad+\frac{2}{3\pi^2}(-6\mu^2+6\mu+6-\pi^2)e^{-2\mu}
+\frac{1}{2\pi^2}(-36\mu^2-66\mu+25-6\pi^2)e^{-4\mu}
+{\cal O}(e^{-6\mu}),
\end{align}
one can see that the perturbative part of $\widetilde J^{(0)}(\mu)$ is indeed a cubic polynomial in $\mu$.
The higher order correction of the WKB expansion can be systematically computed by the method of Wigner transformation \cite{MP}, or the semi-classical analysis of the thermodynamic Bethe ansatz (TBA) equation \cite{CM}.
It turns out that the coefficient $C_k$ and $B_{k,0}$ in \eqref{Jpert} do not receive corrections higher than ${\cal O}(k^1)$ and ${\cal O}(k^3)$, respectively.
On the other hand, the constant $A_k$ in \eqref{Jpert} receives corrections of all order in $k$.

We can continue this process to higher and higher $\hbar$ corrections and
reproduce the Taylor expansion of the functions $a_\ell(k)$ and $b_\ell(k)$ in \eqref{afewMB} up to a certain large order of $k$.
The coefficient of the membrane instanton was first obtained in this way \cite{MP,CM}.

\subsection{Exact values of partition function at finite $N$}
\label{exactvalues}
To fully study the instanton expansion, let us start the computations of the exact values of the partition function at finite $N$.
The density matrix is of the form
\begin{align}
\rho(q_1,q_2)=\frac{E(q_1)E(q_2)}{M(q_1)+M(q_2)}.
\label{kernel}
\end{align}
This type of the integration kernel is quite ubiquitous and appears from time to time in modern physics.
In the context of light-cone string field theory \cite{GSB}, the so-called Neumann coefficients, representing the overlap between one string state and two string state, takes this form.
From the integrability viewpoint, this form has a close relation to the TBA equation and was studied in \cite{TW}.

\begin{table}
\begin{align*}
&Z_1(1)=\frac{1}{4},\quad
Z_1(2)=\frac{1}{16\pi},\quad
Z_1(3)=\frac{-3+\pi}{64\pi},\quad
Z_1(4)=\frac{10-\pi^2}{1024\pi^2},\\&\quad 
Z_1(5)=\frac{26+20\pi-9\pi^2}{4096\pi^2},\quad
Z_1(6)=\frac{78-121\pi^2+36\pi^3}{147456\pi^3},\\
&Z_2(1)=\frac{1}{8},\quad
Z_2(2)=\frac{1}{32\pi^2},\quad
Z_2(3)=\frac{10-\pi^2}{512\pi^2},\quad
Z_2(4)=\frac{24-32\pi^2+3\pi^4}{49152\pi^4},\\
&Z_3(1)=\frac{1}{12},\quad
Z_3(2)=\frac{-3+\pi}{48\pi},\quad
Z_3(3)=\frac{9+108\pi-64\sqrt{3}\pi}{5184\pi},\\
&Z_4(1)=\frac{1}{16},\quad
Z_4(2)=\frac{-8+\pi^2}{512\pi^2},\quad
Z_4(3)=\frac{-8-32\pi+11\pi^2}{8192\pi^2},\\
&Z_6(1)=\frac{1}{24},\quad
Z_6(2)=\frac{54-5\pi^2}{5184\pi^2},\quad
Z_6(3)=\frac{189+192\sqrt{3}\pi-125\pi^2}{186624\pi^2}.
\end{align*}
\caption{The first few exact values of the partition function of the ABJM matrix model. See \cite{HMO2} for more exact values.}
\label{tab:exactZ}
\end{table}

Among others, this expression of the density matrix implies that we can compute the powers of the density matrix without much difficulty.
To explain the computation, we first rewrite \eqref{kernel} schematically as
\begin{align}
\{M,\rho\}=E\otimes E,
\end{align}
if we regard $\rho$, $M$ and $E$ as a symmetric matrix, a diagonal matrix and a vector with continuous indices respectively.
Then, by studying the commutator $[M,\rho^n]$ for even $n$ or the anti-commutator $\{M,\rho^n\}$ for odd $n$, we find
\begin{align}
\rho^n(q_1,q_2)
=\sum_{m=0}^{n-1}(-1)^m
\frac{(\rho^mE)(q_1)(\rho^{n-1-m}E)(q_2)}{M(q_1)-(-1)^nM(q_2)}.
\end{align}
Note that, this formula shows that, instead of multipling the matrices directly, we can pick up a vector $E$ and multiply $\rho$ to it one after another, which is much easier.

Using this method, in \cite{HMO2} we obtained the exact values of the partition function up to $(k,N_\text{max})=(1,44),(2,20),(3,18),(4,16),(6,14)$.
See table \ref{tab:exactZ} for a few examples.
By now with some small technical progress, we believe that we can proceed to more exact values.
See \cite{HO} for many exact values of the ABJ partition function.

\subsection{Grand potential}
After obtaining many exact values, we can now proceed to studying the exact coefficient of the total non-perturbative effects \cite{HMO2}.
The results are given in table \ref{nonpert}.
Since our analysis contains some guesswork, we shall spell out the full details, so that the reader can judge the accuracy by herself.
\begin{table}
\begin{align*}
J_1^\text{np}&=\biggl[\frac{4\mu^2+\mu+1/4}{\pi^2}\biggr]e^{-4\mu}
+\biggl[-\frac{52\mu^2+\mu/2+9/16}{2\pi^2}+2\biggr]e^{-8\mu}
+{\cal O}(e^{-12\mu}),\\
J_2^\text{np}&=\biggl[\frac{4\mu^2+2\mu+1}{\pi^2}\biggr]e^{-2\mu}
+\biggl[-\frac{52\mu^2+\mu+9/4}{2\pi^2}+2\biggr]e^{-4\mu}
+{\cal O}(e^{-6\mu}),\\
J_3^\text{np}&=\frac{4}{3}e^{-\frac{4}{3}\mu}
-2e^{-\frac{8}{3}\mu}
+\biggl[\frac{4\mu^2+\mu+1/4}{3\pi^2}+\frac{20}{9}\biggr]e^{-4\mu}
-\frac{88}{9}e^{-\frac{16}{3}\mu}
+{\cal O}(e^{-\frac{20}{3}\mu}),\\
J_4^\text{np}&=e^{-\mu}
+\biggl[-\frac{4\mu^2+2\mu+1}{2\pi^2}\biggr]e^{-2\mu}
+\frac{16}{3}e^{-3\mu}
+\biggl[-\frac{52\mu^2+\mu+9/4}{4\pi^2}+2\biggr]e^{-4\mu}
+{\cal O}(e^{-5\mu}),\\
J_6^\text{np}&=\frac{4}{3}e^{-\frac{2}{3}\mu}-2e^{-\frac{4}{3}\mu}
+\biggl[\frac{4\mu^2+2\mu+1}{3\pi^2}+\frac{20}{9}\biggr]e^{-2\mu}-\frac{88}{9}e^{-\frac{8}{3}\mu}+{\cal O}(e^{-\frac{10}{3}\mu}).
\end{align*}
\caption{The first few non-perturbative terms of the grand potential of the ABJM matrix model obtained from the numerical study. See \cite{HMO2,HMO3} for more terms.}
\label{nonpert}
\end{table}

Let us pick up the case of $k=4$ for concreteness.
Suppose that we already know the expansion structure of the non-perturbative effects
\begin{align}
J_4^\text{np}(\mu)
=\gamma_1e^{-\mu}+(\alpha_2\mu^2+\beta_2\mu+\gamma_2)e^{-2\mu}
+\gamma_3e^{-3\mu}+\cdots.
\label{J4}
\end{align}
By rewriting the large $\mu$ expansion of the grand potential in terms of the large $N$ expansion of the partition function, we find
\begin{align}
Z_4(N)=Z_4^{(0)}(N)+Z_4^{(1)}(N)+Z_4^{(2)}(N)
+\cdots,
\end{align}
where each term is given by
\begin{align}
Z_4^{(0)}(N)&=e^{A_4}C_4^{-\frac{1}{3}}
\Ai\Bigl[C_4^{-\frac{1}{3}}(N-B_4)\Bigr],\nonumber\\
Z_4^{(1)}(N)&=\gamma_1e^{A_4}C_4^{-\frac{1}{3}}
\Ai\Bigl[C_4^{-\frac{1}{3}}(N+1-B_4)\Bigr],\nonumber\\
Z_4^{(2)}(N)&=\biggl(\alpha_2\partial_N^2-\beta_2\partial_N+\gamma_2+\frac{\gamma_1^2}{2}\biggr)
e^{A_4}C_4^{-\frac{1}{3}}
\Ai\Bigl[C_4^{-\frac{1}{3}}(N+2-B_4)\Bigr],
\end{align}
with $B_4=B_{4,0}$ \eqref{CB}.
Note that the argument of the first non-perturbative effect $Z_4^{(1)}(N)$ is shifted by $1$ because of the exponent factor $e^{-\mu}$ in $J_4^\text{np}(\mu)$ \eqref{J4}.
In the second non-perturbative effect $Z_4^{(2)}(N)$, instead of the polynomial of $\mu$, we replace $\mu$ by $-\partial_N$.
Note also that the lower non-perturbative effects come in from the expansion of the exponential function in the grand canonical partition function.

\begin{table}
\begin{verbatim}
Z={1/16,1/512-1/(64*Pi^2),11/8192-1/(1024*Pi^2)-1/(256*Pi)};
(*Add as many exact values as possible*)
$MaxExtraPrecision=300;
Xi[w_]:=Exp[(ga1) w+(al2 mu^2+be2 mu+ga2) w^2];
(*Add higher order terms w^3, w^4, ... similarly when necessary*)
k = 4; c[k_]:= 2/(Pi^2 k); b[k_]:= 1/(3 k) + k/24;
a[k_]:= If[EvenQ[k], -(Zeta[3]/(Pi^2 k)) \
  - 2/k Sum[m (k/2 - m) Log[2 Sin[(2 Pi m)/k]], {m,1,k/2-1}], \
  -(Zeta[3]/(8 Pi^2 k)) + k/4 Log[2] \
  - 1/k Sum[(k + (-1)^m (2 m - k))/4 (3 k - (-1)^m (2 m - k))/4 \
  Log[2 Sin[(Pi m)/k]], {m,1,k-1}]];
J0 = c[k]^(-1/3) AiryAi[c[k]^(-1/3) (n-b[k])];
J1 = CoefficientList[CoefficientList[Series[Xi[w],{w,0,2}],w][[2]],mu]. \
  Table[(-1)^m D[c[k]^(-1/3) AiryAi[c[k]^(-1/3) (n + 4/k - b[k])],{n,m}], \
  {m,0,0}] // Simplify;
J2 = CoefficientList[CoefficientList[Series[Xi[w],{w,0,2}],w][[3]],mu]. \
  Table[(-1)^m D[c[k]^(-1/3) AiryAi[c[k]^(-1/3) (n + 8/k - b[k])],{n,m}], \
  {m,0,2}] // Simplify;
(*Define higher instanton terms J3, J4, ... similarly when necessary*)
\end{verbatim}
\caption{Mathematica code for studying the coefficients of non-perturbative effects.}
\label{mathematica}
\end{table}

Now let us fit the exact values against this function form.
We first prepare as many exact values as possible.
In the following analysis we shall utilize all the values with $N_\text{max}=16$ in \cite{HMO2}.
As long as we know the command ``FindFit'' in Mathematica, it is by no means difficult to obtain the fitted result.
Still the reader may find it useful to directly apply the given program.
So we also list the program in table \ref{mathematica}.
Note that, since we are only interested in integral $k$, we have used the expression for $a_\ell(k)$ in \cite{HaOk}.
Then, if we ask Mathematica what is the best fit for $\gamma_1$ by
\begin{verbatim}
FindFit[Log[N[Z,50]],a[k]+Log[Abs[J0+J1]],{ga1},n]
\end{verbatim}
we find
\begin{align*}
\gamma_1=0.98721718103894311608972098476342715916438226064401.
\end{align*}
So we have 2\% accuracy for $\gamma_1$.
We are sure that the reader is not satisfied with it.
For example, it is not possible to distinguish $\pi$ from $\sqrt{10}$ with 2\% accuracy.
But it is not difficult to roughly estimate the errors of this value.
Although we have exact values for the partition function, since we neglect higher orders of the non-perturbative corrections in the grand potential, at each order we always have an error of $e^{-4\mu/k}$.
After plugging in the stationary condition $N=\partial J/\partial\mu=C_k\mu^2$ \eqref{stationary}, we find the errors
\begin{align}
e^{-4\mu/k}=e^{-2\pi\sqrt{2N/k}}\simeq 10^{-6},
\end{align}
with $k=4$ and $N\simeq 10$.
Since we have quadratic coefficients at the next order, the errors are actually much bigger than we expect.
But anyway, we can improve our results by including higher instantons
\begin{verbatim}
FindFit[Log[N[Z,50]],a[k]+Log[Abs[J0+J1+J2]],{ga1,al2,be2,ga2},n]
\end{verbatim}
Now the result is much improved
\begin{align}
\gamma_1=1.00000003159520570311264581986409771275067296902425.
\end{align}
We can continue this fitting to higher and higher orders.
For example, for
\begin{verbatim}
FindFit[Log[N[Z,50]],a[k]+Log[Abs[J0+J1+J2+J3+J4+J5]], \
  {ga1,al2,be2,ga2,ga3,al4,be4,ga4,ga5},n]
\end{verbatim}
we find
\begin{align}
\gamma_1=1.00000000000000000000000012119938655342513138232292.
\end{align}
However, if we go to the sixth non-perturbative corrections, the accuracy suddenly drops
\begin{align}
\gamma_1=1.00000000000000000343882868029802074330254814069423.
\end{align}
The reason is that although we only find roughly 25 digit accuracy, we have already used up the 50 digits in the exact values of partition function.
Therefore, we improve by preparing 100 digits from the beginning
\begin{verbatim}
FindFit[Log[N[Z,100]],a[k]+Log[Abs[J0+J1+J2+J3+J4+J5+J6]], \
  {ga1,al2,be2,ga2,ga3,al4,be4,ga4,ga5,al6,be6,ga6},n]
\end{verbatim}
Then we find
\begin{align}
\gamma_1
=0.999999999999999999999999999999999998548662756071218409269666
\cdots.
\end{align}
When we extend to the eighth non-perturbative corrections, an error message appears again
\begin{verbatim}
Failed to converge to the requested accuracy or precision within 100 iterations.
\end{verbatim}
But we can increase to 150 digits
\begin{verbatim}
FindFit[Log[N[Z,150]],a[k]+Log[Abs[J0+J1+J2+J3+J4+J5+J6+J7+J8]], \
  {ga1,al2,be2,ga2,ga3,al4,be4,ga4,ga5,al6,be6,ga6,ga7,al8,be8,ga8},n]
\end{verbatim}
to avoid the error message and find
\begin{align}
\gamma_1
=0.999999999999999999999999999999999999999999999999999994165937
\cdots.
\end{align}
Note that we should end our fitting at this point.
We have only 16 data and if we fit up to the eighth non-perturbative corrections we use up all the data and it does not make sense to fit with more unknowns.
We believe that we can go a little further to find more data with good machines.
However, at present let's be satisfied with more than 50 digit accuracy.
After we are satisfied with the first coefficient, we can simply plug in the correct value and proceed to finding out coefficients of higher orders.

Here we pick up the $k=4$ case and explain our numerical study and the guesswork in details.
We hope we have convinced the reader with the high accuracy we can reach at this point.

\subsection{Comparison with topological strings}
Note that the exponents of the grand potential obtained in table \ref{nonpert} have exactly the same one as the worldsheet instanton $e^{-\frac{4m}{k}}$ in \eqref{fml}.
Hence we can follow the proposal from \cite{MPtop} to compare the result with the free energy of the topological string theory on local ${\mathbb P}^1\times{\mathbb P}^1$ which describes the worldsheet instantons.
We find that the numerical values of the first instanton for $k=3,4,6$ and the values of the second instanton for $k=6$ match with the prediction in \eqref{afewWS} coming from $F^\text{top}$ in \eqref{Jnp-to-Fref}.
However, if we trust this expression and extrapolate to smaller $k$, we find that the match does not hold any more.
Besides, the expression is divergent at $k=1,2$ for the first instanton and $k=1,2,4$ for the second instanton.
In fact this divergence gives us an important clue to understand the whole instanton expansion.
The exponents of the terms containing divergences are $e^{-2\ell\mu}$, which are nothing but the exponents of the membrane instantons.
Besides, if we look at the behavior of the worldsheet instantons around the divergence, we find
\begin{align}
\lim_{k\to 2}d_1(k)e^{-\frac{4\mu}{k}}
&=\biggl[\frac{4}{\pi^2(k-2)^2}+\frac{4(\mu+1)}{\pi^2(k-2)}+\frac{2\mu^2+2\mu+1}{\pi^2}+\frac{1}{3}\biggr]e^{-2\mu},\nonumber\\
\lim_{k\to 4}d_2(k)e^{-\frac{8\mu}{k}}
&=\biggl[-\frac{8}{\pi^2(k-4)^2}-\frac{4(\mu+1)}{\pi^2(k-4)}
-\frac{2\mu^2+2\mu+1}{2\pi^2}-\frac{7}{6}\biggr]e^{-2\mu},
\end{align}
whose finite part looks already close to the expression of the numerical values.
In this way, we can expect that the membrane is also divergent at these points so that totally the divergences cancel among themselves.
From table \ref{nonpert} we also observe that the divergence of the membrane instanton $e^{-2\ell\mu}$ for an odd integer $\ell$ comes from even integers $k$, while the divergence for an even integer $\ell$ comes from all integers $k$.
Also, the coefficients of the membrane instanton $e^{-2\ell\mu}$ for odd $\ell$ vanish at odd integers $k$.
Analytically continuing to non-integral $k$, from this observation, we can expect that the $\ell$-th membrane instanton is given by a periodic function.
Especially, if we assume that it is expressed by using the trigonometric function $\tan\frac{\pi k\ell}{4}$ and simple rational functions, we can determine the function form of the membrane instantons by combining the results of the WKB expansion.
In fact, this is how \eqref{afewMB} was first determined in \cite{HMO2,CM}.

Note that there are no other exponents $e^{-(\frac{4m}{k}+2\ell)\mu}$ labeled by $(m,\ell)$ in \eqref{fml} which have the same contribution as the first and the second membrane instantons $e^{-2\mu}$ and $e^{-4\mu}$ except the pure worldsheet instantons.
For the higher instanton effects, however, the situation is more difficult.
The reason is that besides the worldsheet instantons and the membrane instantons we need to consider the bound states between them $e^{-(\frac{4m}{k}+2\ell)\mu}$ ($m\ne 0,\ell\ne 0$) \cite{HMO3,CM}.
We do not have clues to them from any systematic expansions because the bound state effects are detected neither from the 't Hooft expansions nor from the WKB expansions. 
The only clue is the pole cancellation mechanism as we have explained in section \ref{cancellation} and used above to fix the coefficients of the membrane instantons.
Luckily, we find that if we assume that the bound state effects $f_{m,\ell}(\mu)$ are expressed in terms of the corresponding worldsheet instantons $d_m(k)$ and the corresponding membrane instantons $a_\ell(k)$ as $(A_{m,\ell}(k)=-2m\pi^2 a_\ell(k))$ \cite{HMO3}
\begin{align}
f_{m,\ell}(\mu)=d_m(k)\sum_{(p_1,\cdots,p_\ell)}
\frac{A_{m,1}^{p_1}(k)A_{m,2}^{p_2}(k)\cdots A_{m,\ell}^{p_\ell}(k)}
{p_1!p_2!\cdots p_\ell!},
\end{align}
with the sum running over all of allowed partitions $(p_1,\cdots,p_\ell)$ of $\ell$ satisfying $p_1+2p_2+\cdots+\ell p_\ell=\ell$, we can observe a general cancellation among the worldsheet instantons, the membrane instantons and their bound states, whose finite results coincide beautifully with the numerical studies.
This observation implies that, if we redefine the chemical potential $\mu$ as
\begin{align}
\mu_\text{eff}=\mu+\frac{1}{C_k}\sum_{\ell=1}^\infty a_\ell(k)e^{-2\ell\mu},
\end{align}
or more explicitly as in \eqref{shift} for integral $k$,
\begin{itemize}
\item the bound state effects are absorbed into the worldsheet instanton effects.
\end{itemize}
Besides, we have a bonus from this redefinition:
\begin{itemize}
\item the quadratic polynomial coefficients of the membrane instantons are reduced into linear polynomials,
\begin{align}
J(\mu_\text{eff})=\frac{C}{3}\mu_\text{eff}^3+B\mu_\text{eff}+A
+\sum_{m=1}^\infty d_m(k)e^{-\frac{4m\mu_\text{eff}}{k}}
+\sum_{\ell=1}^\infty
(\widetilde b_\ell(k)\mu_\text{eff}+\widetilde c_\ell(k))
e^{-2\ell\mu_\text{eff}},
\end{align}
and the coefficients $\widetilde c_\ell(k)$ is expressed in terms of $\widetilde b_\ell(k)$ as
\begin{align}
\widetilde c_\ell(k)
=-k^2\frac{\partial}{\partial k}\frac{\widetilde b_\ell(k)}{2\ell k}.
\end{align}
\end{itemize}

\subsection{Topological string theory}
In the above several subsections, we have seen that the WKB expansion, computation of the partition function and the pole cancellation mechanism can determine all of the worldsheet instantons, the membrane instantons and their bound states order by order.
It is not clear, however, how the coefficient functions of higher orders generally look like.
In this subsection, we shall explain a further relation between the ABJ(M) matrix model and topological string theory and determine the coefficient functions systematically.

We shall sketch a relation between the membrane instanton and the refined topological string theory in the NS limit after recapitulating the relation between the worldsheet instanton and the (unrefined) topological strings.
The large $N$ expansion of the free energy in the ABJ(M) theory is formally equivalent to the free energy of pure Chern-Simons theory on $\mathbb{RP}^3$, which in turn is related to the topological string theory on a non-compact Calabi-Yau manifold, known as the local $\mathbb{P}^1\times\mathbb{P}^1$ \cite{MPtop}.
It turns out that the chemical potential in the ABJ(M) Fermi gas is a natural variable in the large radius expansion on the topological string side.
The K\"{a}hler parameters of two $\mathbb{P}^1$'s in local $\mathbb{P}^1\times \mathbb{P}^1$ are identified with the chemical potential as in \eqref{Kahler-T12}, and they are exchanged under the Seiberg-like duality of the ABJ theory.
Surprisingly, it was found in \cite{HMMO} that the non-perturbative corrections in the ABJ(M) matrix model are completely determined by the refined topological string on the local $\mathbb{P}^1\times\mathbb{P}^1$, and the precise relation is given by \eqref{Jnp-to-Fref}.
The first term in \eqref{Jnp-to-Fref} represents the worldsheet instanton, which corresponds to the standard topological string.
This part of the correspondence \eqref{Jnp-to-Fref} is a usual story for the genus expansion of matrix models, and it can be checked, in principle, to any order of the string coupling $g_s=2/k$.
The non-trivial part of the proposal in \cite{HMMO} is the second term in \eqref{Jnp-to-Fref}, namely, the correspondence between the membrane instantons in the ABJ(M) matrix model and the refined topological string in the NS limit.
Note that in \eqref{Jnp-to-Fref} the pole cancellation mechanism works within every BPS multiplet.
Namely, we can check that the combination of $F_\text{top}$ and $F_\text{NS}$ in \eqref{Jnp-to-Fref} with common BPS indices $N_{j_L,j_R}^{\bf d}$ does not contain any divergences.
Also, we stress that not only the singular part but also the finite part are correctly reproduced from \eqref{Jnp-to-Fref}.

As discussed in \cite{Mironov:2009uv,Mironov:2009dv,Aganagic:2011mi}, the refined topological string theory in the NS limit are closely related to the quantization of mirror curves of local Calabi-Yau manifolds.
One can easily see that the Fermi surface $H(x,p)=0$ of the ABJM Fermi gas corresponds to the mirror curve of the diagonal local $\mathbb{P}^1\times\mathbb{P}^1$.
More generally, the mirror curve of  local $\mathbb{P}^1\times \mathbb{P}^1$ is given by
\begin{align}
H(x,p)= -1+e^x+e^p+z_1 e^{-x}+z_2 e^{-p}=0.
\label{mirror-curve}
\end{align}
We can ``quantize'' this curve with $[x,p]=i\hbar$, where the Planck constant $\hbar$ of this quantum mechanical system is unrelated to the previous Planck constant in the Fermi gas system and is determined later.
The corresponding difference equation for the wave function $\Psi(x)$ is 
\begin{align}
(-1+e^x+z_1 e^{-x})\Psi(x)+\Psi(x+\hbar)+z_2\Psi(x-\hbar)=0,
\label{Sch-eq}
\end{align}
whose formal solution
\begin{align}
\Psi(x)=\exp\left[\frac{1}{\hbar}S(x,\hbar)\right],\quad
S(x,\hbar)=\sum_{n=0}^\infty S_n(x)\hbar^{2n},
\end{align}
gives the ``quantum periods''
\begin{align}
\Pi_A=\oint_A dS,\quad\Pi_B=\oint_B dS.
\end{align}
Using the quantum periods, we can compute the free energy in the NS limit\cite{Aganagic:2011mi}.

After identifying the two moduli $z_I$ $(I=1,2)$ in \eqref{mirror-curve} and $\hbar$ as
\begin{align}
z_I=e^{-\frac{T_I}{g_s}},\quad \hbar=\frac{2\pi i}{g_s}=\pi ik,
\end{align}
we find that the membrane instanton coefficients $a_\ell(k)$ and $b_\ell(k)$ appear as the coefficients of the quantum A-period $\Pi_A$ and the quantum B-period $\Pi_B$, respectively, in the small $z_I$ expansion.
Let us see explicitly how it works in the following.

We shall explain how to compute the quantum periods systematically up to any desired order in the small $z_I$ expansion.
As discussed in \cite{HMMO,Aganagic:2011mi} it is convenient to introduce $V(x)$ by
\begin{align}
V(x)=\frac{\Psi(x+\hbar)}{\Psi(x)}.
\end{align}
Then \eqref{Sch-eq} is rewritten as
\begin{align}
V(X)=1-X-\frac{z_1}{X}-\frac{z_2}{V(Xq^{-1})},
\label{Vx-eq}
\end{align}
where we have also introduced $X=e^x$ and $q=e^\hbar$.
One can easily solve \eqref{Vx-eq} in the small $z_I$ expansion as
\begin{align}
V(X)=1-X-\frac{z_1}{X}+\frac{qz_2}{X-q}
-\frac{q^3z_1z_2}{X (X-q)^2}+\frac{q^4z_2^2}{(X-q)^2 \left(X-q^2\right)}
+\mathcal{O}(z^3).
\end{align}
Once we know $V(X)$, the quantum periods are obtained by rewriting the contour integral as a sum of residues \footnote{Here we will use slightly different normalization of quantum periods as compared to \cite{HMMO}.}
\begin{align}
\Pi_A(z_1,z_2;q)&=-\text{Res}_{X=0} 
\frac{1}{X}\log\left[\frac{V(X)}{1-X}\right],
\nonumber\\
\Pi_B(z_1,z_2;q)&=-\sum_{j\geq 0}\text{Res}_{X=q^j}
\frac{\log(X)}{X}\log\left[\frac{V(X)}{1-X}\right].
\label{Pi-residue}
\end{align}
The quantum A-period is given by
\begin{align}
\Pi_A(z_1,z_2,q)=z_1+z_2+(4+q+q^{-1})z_1z_2+\frac{3}{2}(z_1^2+z_2^2)
+\mathcal{O}(z^3),
\end{align}
which defines a quantum version of the mirror map
\begin{align}
\frac{1}{2}\log \frac{Q_I}{z_I}=\Pi_A(z_1,z_2,q).
\end{align}
Here $Q_I$ is related to the ``effective'' K\"{a}hler parameter $T_I^{\text{eff}}$ by
\begin{align}
Q_I=e^{-\frac{T_I^{\text{eff}}}{g_s}},
\end{align} 
which takes care of the effect of the bound states.
Inverting the quantum mirror map, we can express $z_I$ in terms of $Q_I$
\begin{align}
-\frac{1}{2}\log\frac{z_I}{Q_I}
=Q_1-\frac{Q_1^2}{2}+Q_2-\frac{Q_2^2}{2}
+\frac{q^2-q^{-2}}{q-q^{-1}}Q_1Q_2+\mathcal{O}(Q^3).
\label{Q-mirror-map}
\end{align}
On the other hand, the quantum B-period is related to the derivative of free energy in the NS limit as
\begin{align}
Q_2\frac{\partial}{\partial Q_2}F_\text{NS}(Q_1,Q_2)&=-\frac{2}{\hbar}
\Big[2\Pi_B^\text{even}(z_1,z_2;q)-\Pi_A(z_1,z_2,q)^2\Big],\nonumber\\
Q_1\frac{\partial}{\partial Q_1}F_\text{NS}(Q_1,Q_2)&=-\frac{2}{\hbar}
\Big[2\Pi_B^\text{even}(z_2,z_1;q)-\Pi_A(z_1,z_2,q)^2\Big],
\label{deriv-FNS}
\end{align}
where $\Pi_B^\text{even}$ denotes the even-power part of the quantum B-period in the $\hbar$ expansion.
Calculating the residue in \eqref{Pi-residue}, the right-hand-side of \eqref{deriv-FNS} is easily found to be
\begin{align}
\frac{1}{\hbar}\Big[2\Pi_B^\text{even}(z_1,z_2;q)-\Pi_A(z_1,z_2,q)^2\Big]=\frac{q+1}{q-1}z_2+\frac{(q+1)^3}{(q-1)q}z_1z_2+\frac{5+8q+5q^2}{2(q^2-1)}z_2^2+\mathcal{O}(z^3).
\end{align}
Rewriting the above expression in terms of $Q_I$ using the quantum mirror map \eqref{Q-mirror-map} and integrating \eqref{deriv-FNS}, finally we arrive at
\begin{align}
F_\text{NS}(Q_1,Q_2,q)
=-\frac{q+1}{q-1}(Q_1+Q_2)-\frac{(q^2+1)}{4(q^2-1)}(Q_1^2+Q_2^2)
-\frac{(q^2+1)(q+1)}{q(q-1)}Q_1Q_2+\mathcal{O}(Q^3).
\end{align}
Plugging this expression into \eqref{Jnp-to-Fref}, we can show that our formula \eqref{Jnp-to-Fref} reproduces the all known results of the membrane instantons and the bound states in the ABJ(M) matrix model \cite{HMMO,HO}.

From this non-trivial correspondence in the ABJ(M) theory between the membrane instantons and the refined topological string theory in the NS limit, in \cite{HMMO} it is further conjectured that this correspondence holds for general local Calabi-Yau manifolds, and the combination of $F_\text{top}$ and $F_\text{NS}$ appearing in \eqref{Jnp-to-Fref} gives a non-perturbative completion of the topological string theory.
However, there is one subtle point: we have to turn on an extra discrete $B$-field along the worldsheet instanton part in order for the pole cancellation to work for the general case.
See \cite{HMMO} for the detail.

In \cite{KM,K,HW,GHM1,WZH}, it is argued that this correspondence between the membrane instantons and the refined topological string theory in the NS limit in the ABJ(M) theory can be naturally understood from the ``dual'' perspective, by studying the spectral problem \eqref{QM} of the Hamiltonian in the Fermi gas system.

\section{Related topics} \label{topics}
In the previous section, we have reviewed the progress in understanding the partition functions of the ABJM theory and the ABJ theory.
Here we shall explain some progress after finding the exact instanton expansion of the ABJ(M) partition function.

\subsection{Exact results with supersymmetry enhancement}\label{N=8}
As seen before, the large $\mu$ expansion of the grand potential is completely determined by the refined topological string theory on local $\mathbb{P}^1 \times \mathbb{P}^1$ \cite{HMMO}.
However, it is not technically easy to compute the higher order instanton corrections in practice.
It is well-known that at $k=1,2$, the supersymmetry of the ABJM theory is enhanced from $\mathcal{N}=6$ to $\mathcal{N}=8$.
It is natural to expect that for these levels some simplifications happen.
In fact, as we will see below, the large $\mu$ expansion of the grand potential drastically simplifies at $k=1,2$ \cite{CGM}.
We can write it down in a closed form in terms of the topological string free energy.

Let us first write down the results in \cite{CGM}.
The complete large $\mu$ expansion of $J_k(\mu)$ at $k=1,2$ is exactly given by
\begin{align}
J_1(\mu)&=\frac{1}{16\pi^2}\(F_0(t)-t F_0'(t)+\frac{1}{2}t^2F_0''(t)\)
+\frac{3\mu}{8}+\frac{\log 2}{4}+F_1(t)+F_1^\text{NS}(t),\notag \\
J_2(\mu)&=\frac{1}{4\pi^2}\(F_0(t)-t F_0'(t)+\frac{1}{2}t^2F_0''(t)\)
+\frac{\mu}{4}+F_1(t)+F_1^\text{NS}(t).
\label{eq:J12-exact}
\end{align}
We need to explain the notations in these equations.\footnote{
We denote the effective K\"ahler modulus $T^\text{eff}$ by $t$ in the current case with fixed $k$.
Since there are two K\"ahler moduli $t_1$ and $t_2$ in local $\mathbb{P}^1 \times \mathbb{P}^1$, the free energy is in general a function of these two parameters $(t_1,t_2)$.
Here we denote the free energy in the diagonal slice by $F_g(t,t)=F_g(t)$.
}
The functions $F_0(t)$ and $F_1(t)$ are the standard genus zero and genus one free energies on local $\mathbb{P}^1 \times \mathbb{P}^1$, respectively.
These are computed in a standard way of the special geometry.
The function $F_1^\text{NS}(t)$ is the first correction to the refined topological string free energy in the NS limit.
The K\"ahler modulus $t$ is related to the complex modulus $z$ by the mirror map
\begin{align}
t=-\log z+4z\, {}_4F_3\(1,1,\frac{3}{2},\frac{3}{2};2,2,2;-16z\). 
\label{eq:mirrormap}
\end{align}
The only difference between $k=1$ and $k=2$ is the identification of $z$ and $\mu$:
\begin{align}
z=e^{-4\mu} \;\;\mbox{ for }\;\; k=1, \quad
z=e^{-2\mu} \;\;\mbox{ for }\;\; k=2.
\end{align}
At large radius point ($t \to \infty$), the genus zero free energy is expanded as
\begin{align}
F_0(t)=\frac{t^3}{6}-2\zeta(3)+4e^{-t}-\frac{9}{2}e^{-2t}+\cO(e^{-3t}).
\end{align}
Eliminating $t$ by \eqref{eq:mirrormap}, one easily finds
\begin{align}
&F_0(t)-t F_0'(t)+\frac{1}{2}t^2F_0''(t)
=-\frac{1}{6}\log^3 z-2\zeta(3)\notag\\&\qquad
+4(\log^2 z-\log z+1)z
-\frac{1}{2}(52\log^2 z-2\log z+9)z^2+\cO(z^3).
\end{align} 
The free energies $F_1(t)$ and $F_1^\text{NS}(t)$ are exactly given by
\begin{align}
F_1(t)&=-\frac{1}{12}\log [ 64z(1+16z) ]
-\frac{1}{2} \log \biggl[ \frac{K(-16z)}{\pi} \biggr]
=-\frac{\log z}{12}+\frac{2}{3}z-\frac{10}{3}z^2+\cO(z^3),
\notag\\
F_1^\text{NS}(t)&=\frac{1}{12}\log z-\frac{1}{24}\log(1+16z) 
=\frac{\log z}{12}-\frac{2}{3}z+\frac{16}{3}z^2+\cO(z^3).
\end{align}
where $K(z)$ is the complete elliptic integral of the first kind.
Plugging these results into \eqref{eq:J12-exact}, one can check that the large $\mu$ expansion in table~\ref{nonpert} is correctly reproduced.

In the following, we briefly explain how to obtain these results.
Here we focus on the case of $k=2$.
What we should do is to take the limit $k \to 2$ in the general formula \eqref{Jnp-to-Fref}.
As mentioned before, both the worldsheet instanton correction and the membrane instanton correction are divergent in this limit, but the sum of them is finite.
Since we already know the divergences are canceled, we can focus on the finite parts.
Let us see the worldsheet instanton part in \eqref{Jnp-to-Fref}.
It is easy to see that the finite part comes only from $g=0,1$.
The $g=1$ part is trivial.
In the limit $k \to 2$, each term of the $g=0$ part is
\begin{align}
\lim_{k \to 2}\(2 \sin \frac{2\pi w}{k} \)^{-2} e^{-\frac{4dw\mu}{k}}
\sim \frac{e^{-2d w \mu}}{12\pi^2 w^2}(3+\pi^2w^2+6dw\mu+6d^2w^2\mu^2).
\end{align}
where $\sim$ means that we extract only the finite part. 
Therefore the finite part of the worldsheet instanton correction is finally given by
\begin{align}
\frac{1}{4\pi^2}F_0^\text{inst}(t)-\frac{t}{4\pi^2}\pd_t F_0^\text{inst}(t)+\frac{t^2}{8\pi^2}\pd_t^2 F_0^\text{inst}(t)+F_1^\text{inst}(t),
\end{align}
where $F_g^\text{inst}(t)$ is the instanton part in $F_g(t)$.
We have used the relation $t=2\mu_\text{eff}$ at $k=2$.
Similarly, as shown in \cite{CGM}, the finite part of the membrane instanton correction is finally given by
\begin{align}
\frac{1}{8}(\pd_{t_1}-\pd_{t_2})^2F_0(t_1,t_2)|_{t_1=t_2=t}+F_1^\text{NS,inst}(t)=-\frac{t}{8}+\frac{\mu}{4}+F_1^\text{NS,inst}(t).
\end{align}
From these results, one obtains the exact expressions \eqref{eq:J12-exact}.

Once the modified grand potential is known, one can compute the grand partition function.
As seen before, the grand potential is constructed from $J_k(\mu)$ by \eqref{grandpot}, with the subscript $M$ dropped here.
Let us plug the result \eqref{eq:J12-exact} into \eqref{grandpot}.
For $k=2$, using
\begin{align}
e^{J_2(\mu+2\pi i n)}=e^{J_2(\mu)}\exp \left[ \pi i n^2 \tau+2\pi i n\( \xi-\frac{1}{12} \) \right],
\end{align}
with
\begin{align}
\tau=\frac{2i}{\pi} F_0''(t),\quad
\xi=\frac{1}{2\pi^2}(tF_0''(t)-F_0'(t)),
\end{align}
we find that the grand partition function is expressed in terms of the elliptic theta function by
\begin{align}
\Xi_2(\mu)=e^{J_2(\mu)} \vartheta_3 \( \xi-\frac{1}{12},\tau \).
\label{eq:Xi2}
\end{align}
Similarly, the grand partition function at $k=1$ is given by
\begin{align}
\Xi_1(\mu)=e^{J_1(\mu)} \vartheta_3 \( \frac{\xi}{2}-\frac{7}{24},\tau \).
\label{eq:Xi1}
\end{align}
We stress that these expressions are \textit{exact}.
All the instanton corrections are encoded through the topological string free energies $F_0(t)$, $F_1(t)$ and $F_1^\text{NS}(t)$.

There are two important consequences of the exact expressions \eqref{eq:Xi2} and \eqref{eq:Xi1}.
One is that one can know the eigenvalues of the density matrix $\rho$ as the zeros of $\Xi_k(\mu)$.
In fact, it is well-known that the theta function has an infinite number of zeros.
It was confirmed in \cite{CGM} that the zeros in \eqref{eq:Xi2} and \eqref{eq:Xi1} precisely reproduce the eigenvalues of ${\rho}$ at $k=1,2$, numerically computed by the exact quantization condition in \cite{KM}.
The other important point is that it is possible to analytically continue $\Xi_k(\mu)$ from $\mu \to \infty$ to $\mu \to -\infty$.
The continuation is essentially the modular S-transformation $\tau \to \bar{\tau}=-1/\tau$.
In \cite{CGM}, this analytic continuation was indeed performed, and then the small $e^{\mu}$ expansion was computed.
The results perfectly agree with the exact values of the partition function in table~\ref{tab:exactZ} (see \cite{HMO1,HMO2} for more values)!

Furthermore, in \cite{CGM,GHM2} the exact results for ABJM at $k=4$ and ABJ at $(k,M)=(2,1)$ were presented.
These also reproduce the known results perfectly.
Note that at $k=4$ the supersymmetry is no longer enhanced.
Nevertheless, we can write down the grand potential in a closed form.
Moreover, in \cite{GHM2}, exact functional equations among the grand partition functions in the ABJ matrix model were found.
These relations are quite similar to so-called quantum Wronskian relations \cite{BLZ2}.
It seems to imply an interesting connection to integrable models. 

\subsection{'t Hooft expansion and Borel resummation}
Throughout this paper, we focus on the large $N$ expansion in the M-theory limit \eqref{Mregime}.
As already seen in subsection \ref{history}, it is also interesting to consider the 't Hooft limit \eqref{tHooftregime}.
In this limit, the ABJM theory is dual to type IIA string theory on $AdS_4 \times \mathbb{CP}^3$,
and the large $N$ expansion is related to the genus expansion in this string theory.
In the 't Hooft limit, the free energy admits the large $N$ expansion \eqref{eq:genus}.
The genus $g$ free energy $F_g(\lambda)$ can be exactly computed by using the holomorphic anomaly equation order by order \cite{DMP1}.
Then, the genus expansion \eqref{eq:genus} turned out to be an asymptotic series for given finite $\lambda$.
The standard way to resum such formal asymptotic series is known as Borel resummations.
For the genus expansion \eqref{eq:genus}, let us define its Borel transform by
\begin{align}
\mathcal{B}F(\zeta)=\sum_{g=2}^\infty \frac{F_g(\lambda)}{(2g-3)!} \zeta^{2g-2}.
\label{eq:Borel-trans}
\end{align}
Remarkably, as observed in \cite{DMP2, GMZ}, this Borel transform has no singularities on the positive real axis.
Therefore the following Borel summation is well-defined:
\begin{align}
\mathcal{S} F(N,\lambda)=N^{-2}F_0(\lambda)+F_1(\lambda)+\int_0^\infty \frac{d\zeta}{\zeta} e^{-\zeta}
\mathcal{B}F(N \zeta).
\label{eq:Borel-resum}
\end{align}
The crucial conclusion in \cite{GMZ} is that this Borel summation does not reproduce the exact values of $F(N,\lambda)$.
There is an exponentially suppressed correction between them:
\begin{align}
F^\text{exact}(N,\lambda)-\mathcal{S} F(N,\lambda)=\mathcal{O}(e^{-\pi N\sqrt{2/\lambda}}).
\end{align}
This correction is nothing but the membrane instanton correction!
It is not captured by the perturbative $1/N$ expansion.
Thus in the 't Hooft limit, the perturbative genus expansion and its Borel resummation are insufficient to reconstruct the exact free energy.
We need to take into account the membrane instanton corrections in addition to the perturbative expansion.
It is an open problem how to incorporate such membrane instantons in the Borel analysis.
As pointed out in \cite{DMP2}, such corrections should be related to complex singularities of the Borel transform \eqref{eq:Borel-trans}.

As reviewed in this paper, we already know the complete M-theoretic expansion of the free energy with the help of the topological string.
It should be in principle possible to use this result to understand the membrane instanton corrections to the genus expansion.
Interestingly, it was observed in \cite{HO2} that the Borel resummation \eqref{eq:Borel-resum} already contains a membrane instanton-like correction in the M-theoretic expansion from the Fermi gas result.
It would be interesting to reveal a quantitative relation between the 't Hooft expansion and the M-theoretic expansion.

\subsection{Other theories}\label{generalization}

In section \ref{N=8} we have seen special cases with supersymmetry enhancements.
Here, let us see how the analysis can be generalized to other theories of the M2-branes.
Especially, when we are interested in gaining the universal property of the M2-branes, it is very important to generalize to as many theories as possible.
We shall list some results obtained recently.

First, let us give a viewpoint from the quiver diagrams, where the vertices represent the gauge group factor $U(N)$ and the edges represent the bi-fundamental matters between the two factors connected.
Using the quiver diagram, the ABJM matrix model is nothing but the Dynkin diagram of affine $A_1$ or $\widehat A_1$.
Then, it is natural to ask how about a general matrix model associated with $\widehat A_r$.

It was already found in \cite{MP} that if the density matrix is hermitian, the perturbative part of any $\widehat A_r$ matrix model is summed up to the Airy function with the coefficient $C$ given explicitly in \cite{GHP,MP}.
The other coefficients are not easy to obtain in general.
In \cite{HoMo} it was found that as long as the quiver diagram associated with the levels is a repetition of an original quiver diagram, we can express the grand potential of the new quiver diagram completely in terms of that of the original diagram.
In \cite{MN1} it was restricted to the ${\cal N}=4$ case \cite{IK4} where the Chern-Simons levels are given by
\begin{align}
k_a=\frac{k}{2}(s_a-s_{a-1}),\quad s_a=\pm 1,
\end{align}
and the explicit form of the coefficient $B$ was found to be
\begin{align}
B&=\frac{B^{(0)}}{k}+kB^{(2)},
\label{B}
\end{align}
with
\begin{align}
B^{(0)}&=-\frac{1}{6}\biggl[
\frac{\Sigma(p)}{\Sigma(q)}+\frac{\Sigma(q)}{\Sigma(p)}
-\frac{4}{\Sigma(q)\Sigma(p)}\biggr],\nonumber\\
B^{(2)}&=\frac{1}{24}\biggl[\Sigma(q)\Sigma(p)
-12\biggl(\frac{\Sigma(q,p,q)}{\Sigma(q)}
+\frac{\Sigma(p,q,p)}{\Sigma(p)}
-\frac{\Sigma(q,p)\Sigma(p,q)}{\Sigma(q)\Sigma(p)}\biggr)\biggr].
\label{B0B2}
\end{align}
Here we adopt the notation of $\Sigma(L)$, with $L$ denoting an alternating sequence of $q$ and $p$, whose definition is given by
\begin{align}
&\Sigma(q)=\sum_{a=1}^mq_a,\quad
\Sigma(p)=\sum_{a=1}^mp_a,\nonumber\\
&\Sigma(q,p)=\sum_{1\le a\le b\le m}q_ap_b,\quad
\Sigma(p,q)=\sum_{1\le a<b\le m}p_aq_b,\nonumber\\
&\Sigma(q,p,q)=\sum_{1\le a\le b<c\le m}q_ap_bq_c,\quad
\Sigma(p,q,p)=\sum_{1\le a<b\le c\le m}p_aq_bp_c.\label{Sigmadef}
\end{align}
Note that the condition in each sum can be stated as the requirement that we choose $q_a$ and $p_a$ out of a sequence $q_1,p_1,q_2,p_2,\cdots,q_m,p_m$ by respecting its ordering.

Furthermore, if we restrict ourselves to the special ${\cal N}=4$ $(q,p)$ model where the $q$ edges with $s_a=+1$ and the $p$ edges with $s_a=-1$ are separated, we find that the coefficient $A$ is given in terms of that of the ABJM matrix model by \cite{MN1}
\begin{align}
A=\frac{1}{2}\bigl(p^2A_{\rm ABJM}(qk)+q^2A_{\rm ABJM}(pk)\bigr).
\label{A}
\end{align}
A similar proposal was made in \cite{HaOk} for the so-called M-theoretic matrix model \cite{MePu,GM}, whose Fermi gas Hamiltonian is related to that of the $(q,1)$ model by a canonical transformation \cite{ADF}.
In \cite{MN2}, the membrane instanton of the $(q,p)$ model was studied and a systematic expansion in terms of the derivative operators acting on the classical expression was found.
Using the results we were able to study the $(2,2)$ model in full details \cite{MN3} and find that the non-perturbative effects are given by the refined topological string theory as in the ABJM matrix model \eqref{pertnp}.
It still awaits to identify the Calabi-Yau manifold whose BPS index describes the $(2,2)$ model, but we observed that the diagonal Gopakumar-Vafa invariant is nothing but that of the $D_5$ del Pezzo surface.
It is interesting to find that the correspondence between the topological string theory and the matrix model coming from the superconformal theory, is not restricted to the case of the ABJM theory.
In \cite{HHO}, the non-perturbative effects were further explored.
Based on the systematic WKB analysis and exact computations in various special cases, analytic expressions of the first few membrane instanton corrections and worldsheet instanton corrections were conjectured for general $(q,p)$ and $k$.

The generalization, however, is not restricted to the $\widehat A$ quiver.
In \cite{GAH} an $\widehat A\widehat D\widehat E$ sector was specified by requiring the long range force among the eigenvalues vanishes.
It is then interesting to ask whether there is a Fermi gas formalism for these series and whether the perturbative corrections are still summed up to the Airy function.
In \cite{DF,MN4} we continue to study the $\widehat D$ quiver.
We find that after using the Cauchy determinant \eqref{Cauchy} in a different way we can still give a Fermi gas formalism to this type of theories and prove the previous proposal \cite{CHJ}
\begin{align}
C=\frac{1}{\pi^2k}
\biggl(\frac{1}{\sigma_0\sigma_1}
+\sum_{m=1}^{r-1}\frac{s_m-s_{m+1}}{\sigma_m\sigma_{m+1}}
+\frac{s_r}{\sigma_r\sigma_{r+1}}\biggr),
\label{C}
\end{align}
with the variables $\sigma_m$ given as
\begin{align}
\sigma_m=\sum_{n=1}^r(|s_m-s_n|+|s_m+s_n|)-4|s_m|,\quad
\sigma_0=2(r-2),\quad\sigma_{r+1}=2\sum_{n=1}^r|s_n|,
\label{sigma&s}
\end{align}
by the reordered $s$,
\begin{align}
0\le |s_r|\le |s_{r-1}|\le \cdots \le |s_1|.
\label{sorder}
\end{align}
As in the ${\cal N}=4$ case of the $\widehat A$ quiver, here we can also restrict ourselves to the case of uniform $s_a$, $s_a=1$, and find that the coefficient $A$ is consistent with
\begin{align}
A=\frac{1}{2}
\bigl(A_{\text{ABJM}}(2rk)+r^2A_{\text{ABJM}}(2(r-2)k)\bigr),
\end{align}
up to ${\cal O}(k^5)$, which is very similar to \eqref{A} for the $(q,p)$ model.
We hope that all these findings will finally be completed to a more systematic understanding.

\section{Summary and further directions}
In this review article, we have explained carefully the quantum effects of the ABJM theory and the ABJ theory, where the partition function and the vacuum expectation values of the half-BPS Wilson loop can be studied in full details including both the perturbative and non-perturbative corrections.

We have seen many interesting properties.
For example, the leading large $N$ behavior of the free energy is given by $\sim k^{\frac{1}{2}}N^{\frac{3}{2}}$.
The perturbative $1/N$ corrections are summed up to an Airy function.
The non-perturbative corrections include the effects interpreted as worldsheet instantons and those interpreted as membrane instantons.
In general, the coefficients of both instantons and their bound states can be divergent at certain values of $k$, though as a whole the divergences in these instanton effects cancel among themselves and the final result is completely finite.

Also, in section \ref{generalization} we have seen that many generalizations enjoy these interesting properties.
Hence, one further direction is to generalize our analysis to as many theories as possible and identify this class clearly.
It is surprising to find that the superconformal Chern-Simons theory of the $\widehat D$ quiver can also be formulated with the Fermi gas system.
It is then natural to ask whether we can apply the same method even for the $\widehat E$ quiver or their cousins with the orthogonal groups or the symplectic groups \cite{GHN,MePu}.
We hope that through these studies we will end up with a universal understanding of the M2-branes.

Next, let us also stress that, after a large number of numerical analysis, our large $\mu$ expansion of the grand potential (or large $N$ expansion of the partition function) remains to be a conjecture.
It is of course interesting to see how this conjecture is proved from a mathematical method.
The study explained in section \ref{N=8} would be a first attempt in this direction.

Finally, let us come back to a more physical question.
Our original motivation in studying the ABJM partition function is to understand the M2-branes.
After seen all of the perturbative and non-perturbative corrections, it is reasonable to come back to ask what we have learned on the M2-brane interactions.
Unfortunately, we do not have much to say about it at this stage, except that some of the results can be rederived from the supergravity \cite{AHHS,BGMS,DDG}.
However, we believe that the beautiful structure we find would be a guide towards a clearer understanding of the $N^{\frac{3}{2}}$ law or even a better description of the mysterious M-theory.

\section*{Acknowledgements}
We are grateful to all of our collaborators at various stages of this project for valuable discussions, including Hiroyuki Fuji, Alba Grassi, Shinji Hirano, Masazumi Honda, Marcos Marino, Tomoki Nosaka and Masaki Shigemori.
The work of S.M.\ was supported in part by JSPS Grant-in-Aid for Scientific Research (C) \# 26400245.
The work of K.O. was supported in part by JSPS Grant-in-Aid for Young Scientists (B) \# 23740178.

\end{document}